\documentclass[twocolumn]{aastex63}

\usepackage{amsfonts}
\usepackage{fancyhdr}
\usepackage{amsmath}
\usepackage{ragged2e}
\usepackage{enumitem}
\usepackage[version=4]{mhchem}
\usepackage{natbib}
\usepackage{multirow}
\definecolor{pink}{rgb}{1,0.1,.6}

\shorttitle{The transmission spectrum of the potentially rocky planet L~98-59~c}
\shortauthors{Barclay et al.}
\graphicspath{{./}{figures/}}

\begin{document}

\title{The transmission spectrum of the potentially rocky planet L~98-59~c}

\correspondingauthor{Thomas Barclay}
\email{thomas.barclay@nasa.gov}

\author[0000-0001-7139-2724]{Thomas Barclay}
\affiliation{NASA Goddard Space Flight Center, 8800 Greenbelt Road, Greenbelt, MD 20771, USA}

\author{Kyle B. Sheppard}
\affiliation{University of Maryland, College Park, MD 20742, USA}

\author[0000-0001-8079-1882]{Natasha Latouf}
\affiliation{NASA Goddard Space Flight Center, 8800 Greenbelt Road, Greenbelt, MD 20771, USA}
\affiliation{George Mason University, 4400 University Dr, Fairfax, VA 22030, USA}

\author{Avi M. Mandell}
\affiliation{NASA Goddard Space Flight Center, 8800 Greenbelt Road, Greenbelt, MD 20771, USA}

\author[0000-0003-1309-2904]{Elisa V. Quintana}
\affiliation{NASA Goddard Space Flight Center, 8800 Greenbelt Road, Greenbelt, MD 20771, USA}

\author[0000-0002-0388-8004]{Emily A. Gilbert}
\affiliation{Jet Propulsion Laboratory, California Institute of Technology, Pasadena, CA 91109, USA}

\author{Giuliano Liuzzi}
\affiliation{NASA Goddard Space Flight Center, 8800 Greenbelt Road, Greenbelt, MD 20771, USA}
\affiliation{University of Basilicata, Via dell'Ateneo Lucano 10, Potenza (PZ), 85100, Italy}

\author[0000-0002-2662-5776]{Geronimo L. Villanueva}
\affiliation{NASA Goddard Space Flight Center, 8800 Greenbelt Road, Greenbelt, MD 20771, USA}

\author[0000-0001-8020-7121]{Giada Arney}
\affiliation{NASA Goddard Space Flight Center, 8800 Greenbelt Road, Greenbelt, MD 20771, USA}

\author{Jonathan Brande}
\affiliation{Department of Physics and Astronomy, University of Kansas, 1082 Malott, 1251 Wescoe Hall Drive, Lawrence, KS 66045, USA}

\author[0000-0001-8020-7121]{Knicole D. Col\'on}
\affiliation{NASA Goddard Space Flight Center, 8800 Greenbelt Road, Greenbelt, MD 20771, USA}

\author[0000-0002-2553-096X]{Giovanni Covone}
\affiliation{Department of Physics "Ettore Pancini", Università di Napoli Federico II, Napoli, Italy}
\affiliation{INAF - Osservatorio Astronomico di Capodimonte, via Moiariello 16, 80131 Napoli, Italy}

\author{Ian J.M. Crossfield}
\affiliation{Department of Physics and Astronomy, University of Kansas, 1082 Malott, 1251 Wescoe Hall Drive, Lawrence, KS 66045, USA}

\author[0000-0002-0388-8004]{Mario Damiano}
\affiliation{Jet Propulsion Laboratory, California Institute of Technology, Pasadena, CA 91109, USA}

\author[0000-0003-0354-9325]{Shawn D. Domagal-Goldman}
\affiliation{NASA Goddard Space Flight Center, 8800 Greenbelt Road, Greenbelt, MD 20771, USA}

\author[0000-0002-5967-9631]{Thomas J. Fauchez}
\affiliation{NASA Goddard Space Flight Center, 8800 Greenbelt Road, Greenbelt, MD 20771, USA}
\affiliation{Integrated Space Science and Technology Institute, Department of Physics, American University, Washington DC}
% \affiliation{NASA GSFC Sellers Exoplanet Environments Collaboration}

\author{Stefano Fiscale}
\affiliation{Science and Technology Department, Parthenope University of
Naples, CDN IC4, 80143, Naples, Italy}

\author[0000-0002-9138-4788]{Francesco Gallo}
\affiliation{Department of Physics "Ettore Pancini", Università di Napoli Federico II, Napoli, Italy}

\author{Christina L. Hedges}
\affiliation{NASA Goddard Space Flight Center, 8800 Greenbelt Road, Greenbelt, MD 20771, USA}
\affiliation{University of Maryland, Baltimore County, 1000 Hilltop Cir, Baltimore, MD 21250, USA}

\author[0000-0002-0388-8004]{Renyu Hu}
\affiliation{Jet Propulsion Laboratory, California Institute of Technology, Pasadena, CA 91109, USA}

\author[0000-0002-1426-1186]{Edwin S. Kite}
\affiliation{University of Chicago, Chicago, IL 60637, USA}

\author{Daniel Koll}
\affiliation{Peking University, Beijing, China}

\author{Ravi K. Kopparapu}
\affiliation{NASA Goddard Space Flight Center, 8800 Greenbelt Road, Greenbelt, MD 20771, USA}

\author[0000-0001-9786-1031]{Veselin B. Kostov}
\affiliation{NASA Goddard Space Flight Center, 8800 Greenbelt Road, Greenbelt, MD 20771, USA}
\affiliation{SETI Institute, 189 Bernardo Ave, Suite 200, Mountain View, CA 94043, USA}

\author{Laura Kreidberg}
\affiliation{Max Planck Institute for Astronomy, Heidelberg, Germany}

\author{Eric D. Lopez}
\affiliation{NASA Goddard Space Flight Center, 8800 Greenbelt Road, Greenbelt, MD 20771, USA}

\author[0000-0001-5864-9599]{James Mang}
\affiliation{Department of Astronomy, University of Texas at Austin, Austin, TX 78712, USA}

\author[0000-0002-4404-0456]{Caroline V. Morley}
\affiliation{Department of Astronomy, University of Texas at Austin, Austin, TX 78712, USA}

\author{Fergal Mullally}
\affiliation{Constellation Energy, 1310 Point St, Baltimore, MD 21231}

\author[0000-0001-7106-4683]{Susan E. Mullally}
\affiliation{Space Telescope Science Institute, 3700 San Martin Drive, Baltimore, MD, 21218, USA}

\author[0000-0001-9771-7953]{Daria Pidhorodetska}
\affiliation{Department of Earth and Planetary Sciences, University of California, Riverside, CA, USA}

\author[0000-0001-5347-7062]{Joshua E. Schlieder}
\affiliation{NASA Goddard Space Flight Center, 8800 Greenbelt Road, Greenbelt, MD 20771, USA}

\author[0000-0002-5928-2685]{Laura D. Vega}
\affiliation{NASA Goddard Space Flight Center, 8800 Greenbelt Road, Greenbelt, MD 20771, USA}
\affiliation{University of Maryland, College Park, MD 20742, USA}
% \affiliation{Center for Research and Exploration in Space Science \& Technology, NASA/GSFC, Greenbelt, MD 20771, USA}

\author{Allison Youngblood}
\affiliation{NASA Goddard Space Flight Center, 8800 Greenbelt Road, Greenbelt, MD 20771, USA}

\author{Sebastian Zieba}
\affiliation{Max Planck Institute for Astronomy, Heidelberg, Germany}

\begin{abstract}
We present observations of the $1.35\pm0.07$ Earth-radius planet L~98-59~c, collected using Wide Field Camera~3 on the Hubble Space Telescope. L~98-59 is a nearby (10.6 pc), bright (H=7.4 mag), M3V star that harbors three small, transiting planets. As one of the closest known transiting multi-planet systems, L~98-59 offers one of the best opportunities to probe and compare the atmospheres of rocky planets that formed in the same stellar environment. We measured the transmission spectrum of L~98-59~c and the extracted spectrum showed marginal evidence (2.1$\sigma$) for wavelength-dependent transit depth variations that could indicate the presence of an atmosphere. We forward-modeled possible atmospheric compositions of the planet based on the transmission spectrum. Although L~98-59 was previously thought to be a fairly quiet star, we have seen evidence for stellar activity, and therefore we assessed a scenario where the source of the signal originates with inhomogeneities on stellar surface. We also see a correlation between transits of L98-59 c and L98-59 b collected 12.5 hours apart, which is suggestive (but at $<$2$\sigma$ confidence) of a contaminating component from the star impacting the exoplanet spectrum. While intriguing, our results are inconclusive and additional data is needed to verify any atmospheric signal. Fortunately, additional data has been collected from both HST and JWST. Should this result be confirmed with additional data, L~98-59~c would be the first planet smaller than two Earth-radii with a detected atmosphere.
\end{abstract}
%The results indicates a low-density extended atmosphere. The data was best modeled by a Hydrogen-rich atmosphere, with absorption dominated by \chem{H_2O} and \chem{CO_2}. However, other solutions such as an atmosphere rich in \chem{HCN} and \chem{NO_3} describe the data equally well. 

\keywords{Exoplanet atmospheric composition --- Super Earths --- M dwarf stars --- Transmission spectroscopy}

\section{Introduction} \label{sec:intro}
In the post-\emph{Kepler} era, ground and space-based transiting exoplanet searches have focused on detecting small planets orbiting small stars \citep{Demangeon2021, burt2021, newton2021}. The primary reason for favoring small stars, mid M-dwarfs and smaller, is that in the near-term, they are likely to be the only targets where we might feasibly detect an atmosphere around a sub-Neptune-sized planet \citep{Gialluca2021}. Among the first exoplanets to be targeted by the James Webb Space Telescope (JWST) are numerous small planets around cool stars which will enable us, for the first time, to begin to see the diversity of atmospheres on terrestrial worlds \citep{Morley2017,Batalha2018,Batalha2023,LustigYaeger2019}.

Several small planets orbiting low mass stars have been observed using the transmission spectroscopy technique with the Hubble Space Telescope (HST). The most prominent of these are the planets that orbit TRAPPIST-1 \citep{Gillon2016,deWit2016,Gillon2017,dewit2018,wakeford2019b,Garcia2022}. While conclusive atmospheric detection has yet to be made for any of the TRAPPIST-1 worlds, these observations have been informative in producing the first limits on atmospheric density and aerosol properties for these planets. For example, HST observations of the TRAPPIST-1 b and c planets \citep{deWit2016}, as well as HST observations of planets d, e, f, and g in this system \citep{dewit2018}, have ruled out a cloud/haze-free, H$_2$-dominated atmosphere, and have been used to argue that hazy H$_2$-rich atmospheres could explain the HST data \citep{Moran2018}. %Furthermore, these results have motivated JWST observations of the TRAPPIST-1 planets. 
More recently, JWST has observed thermal emission from TRAPPIST-1 b \citep{Greene2023} and c \citep{Zieba2023} that suggest these planets likely have thin to no atmospheres.

In addition to TRAPPIST-1, HST observations have enabled us to put constraints on the atmospheric composition of GJ 1214 b \citep{Kreidberg2014}, GJ 1132 b \citep{swain2021,Mugnai2021,Libby2021}, and HD 97658 b \citep{bourrier2017, Guo2020}, and measure a low-density atmosphere on K2-18 b \citep{benneke2019,Tsiaras2019}, TOI-270 d \citep{MikalEvans2022}, and possibly also 55 Cnc e \citep{bourrier2018, tsiaras2016}. Additionally, \textit{Spitzer} was used to demonstrate spectral signatures on the hot Neptune-sized planet LTT 9779b \citep{Dragomir2020,Crossfield2020} and rule out a thick atmosphere on LHS 3844b \citep{Kreidberg2019}. Ground-based observations have further constrained the atmospheres of GJ 1132 b, LHS 1140 b, and LTT 1445 Ab \citep{DiamondLowe2018,DiamondLowe2020,DiamondLowe2022}.
%%\footnote, and while \citet{swain2021} reported the detection of atmospheric signatures on GJ 1132 b, this has been disputed \citet{Mugnai2021,Libby2021}

Amongst the challenges in detecting and characterizing features in the atmospheres of small, rocky worlds with transmission spectroscopy is that spectral features from the star can contaminate those from the planet. Transmission spectroscopy works by taking the difference between the planet's ``in-transit'' spectrum with the ``out-of-transit'' spectrum. However, the light source filtering through the planet's atmosphere comes from the transit chord. Stellar heterogeneities (in the form of dark, cool spots and bright, hot faculae) that reside on the stellar disk, but are unocculted by a transiting planet (outside of the transit chord), can dilute or enhance the measured transit depth \citep{McCullough2014,Rackham2017,Rackham2018,Apai2018,Espinoza2019,Rackham2019b,Rackham2019,Wakeford2019}. These errors in transit depth propagate to errors in the planet's spectrum, and can ultimately mask or mimic atmospheric features. While stellar spectral contamination has been identified as problematic for more than a decade \citep{Pont2008,Pont2013,Sing2011,Berta2012,McCullough2014,Deming2017}, more recent high-precision observations from JWST have shown it to be a limiting factor in confirming the presence of atmospheres on Earth-size planets \citep{Moran2023}, including those orbiting TRAPPIST-1 \citep{Lim2023}.

 Herein, we present HST spectroscopic observations of a potentially rocky planet, L~98-59~c, with a goal of searching for the presence of an atmosphere. This planet was discovered by the Transiting Exoplanet Survey Satellite (TESS), which was designed to discover small planets orbiting bright, nearby stars \citep{Ricker2015}. A number of TESS discoveries have already been prioritized for JWST observations, and among the most compelling are the planets orbiting L~98-59, a bright ($H$=7.4~mag), nearby (10.6 pc) M3~dwarf. L~98-59 hosts three transiting planets \citep{Kostov2019b}, all of which are smaller than 1.6 R$_\oplus$ with orbital periods shorter than 7.45 days. Additionally, there is a fourth confirmed planet that does not appear to transit, and a candidate fifth planet that is also non-transiting \citep{Demangeon2021}. The three transiting planets (planets b--d) have measured masses of 0.40$\pm$0.14 M$_\oplus$, 2.2$\pm$0.3 M$_\oplus$, and 1.9$\pm$0.3 M$_\oplus$, and radii of 0.85$\pm$0.06 R$_\oplus$, 1.4$\pm$0.1 R$_\oplus$, and 1.5$\pm$0.1 R$_\oplus$ \citep{Kostov2019b,Cloutier2019, Demangeon2021}, confirming the bulk terrestrial composition of L~98-59~b and L~98-59~c and alluding to a significant gaseous envelope for L~98-59~d. The two outer transiting planets (planets c and d) are prime targets for atmosphere characterization because they have some of the highest transmission spectroscopy metric and emission spectroscopy metric \citep{Kempton2018, pidhoro} values of any small planet. L~98-59 provides an excellent opportunity to probe the atmospheres of planets smaller than 1.5 R$_\oplus$ that formed and evolved in the same stellar environment. 

We were awarded 28 orbits on the Wide Field Camera 3 (WFC3) instrument on HST to observe 5 transits of L~98-59~b, and one transit each of planet c and d. No evidence of atmospheric features were seen in the spectrum of L~98-59~b \citep{damiano22}. In this paper we report on the HST observations of L~98-59~c. We modeled the spectrum using two different approaches with a goal of identifying whether or not we could definitively detect and place constraints on an atmospheric signal. We also explored the stellar activity of the star, and compared the observations of L~98-59~c with observations of L~98-59-b that were obtained 12.5 hours apart to look for the presence of stellar contamination that might impact both observations. The properties of L~98-59 and the two inner planets (b and c) are summarized in Table~\ref{tab:l98_lit}. %We found that no strong conclusions are possible with the existing data, and that this planet should continue to be highly prioritized for further investigation.

%, and prospects for future characterization with HST and JWST.

\begin{table*}[ht]
\centering       
\caption{L~98-59 System Stellar and Planetary Properties, adopted from \citet{Demangeon2021}.}\label{tab:l98_lit}
\begin{tabular}{l|lrr}     % 4 columns 
\hline   
     & Parameter & \multicolumn{2}{c}{L~98-59} \\
\hline
    Stellar & Radius [$R_{\odot}$] & \multicolumn{2}{c}{$0.303\pm{0.026}$}\\
    & Mass [$M_{\odot}$] & \multicolumn{2}{c}{$0.273\pm{0.030}$}\\  
    & T$_{\textrm{eff}}$ [K] & \multicolumn{2}{c}{$3415\pm{135}$}\\
    & $\log{g_s}$ [cgs units]  & \multicolumn{2}{c}{$4.86\pm{0.13}$}\\
    & [Fe/H] & \multicolumn{2}{c}{-$0.46\pm{0.26}$}\\
    & Distance [pc] & \multicolumn{2}{c}{$10.6194\pm{0.0032}$}\\
    \hline
     &  & L~98-59~b & L~98-59~c \\
    \hline
    Orbital  & $R_p/R_s$ & $0.02512\pm{0.00072}$ & $0.04088\pm{0.00068}$\\
    & $a/R_s$ & $15.0\pm1.4$ & $19.0\pm1.2$\\
    & $i$ [deg] & $87.7\pm{1.2}$ & $88.1\pm{0.36}$\\
    & $T_c$ [BJD-2457000] & $1366.17067\pm{0.00036}$ & $1367.27375\pm{0.00022}$ \\
    & $\rho$ [g.cm$^{-3}$] & $3.6\pm1.5$ & $4.57\pm{0.85}$ \\
    & $P$ [days] & $2.2531136\pm0.0000015$ & $3.6906777\pm{0.0000026}$ \\
    Planet & $R_p$ [$R_{\oplus}$] & $0.850\pm{0.061}$ & $1.385\pm{0.095}$ \\
    & $M_p$ [$M_{\oplus}$] & $0.40\pm{0.16}$ & $2.22\pm{0.26}$\\
    & $T_{\textrm{eq}, A=0}$ [K] & $627\pm{36}$ & $553\pm{27}$\\
    & $a$ [AU] & $0.02191\pm{0.00084}$ & $0.0304\pm{0.0012}$ \\
\hline
\end{tabular}
\end{table*}
 
\section{Data analysis}
\label{sec:data}
A transmission spectrum of L~98-59~c was measured from a single transit observed with HST/WFC3 with the G141 grism on 7 April 2020 (HST Program GO-15856, PI T. Barclay), a visit lasting for four HST orbits. The observations with the grism were in round trip spatial scanning mode with the GRISM512 subarray, with NSAMP=4 and the SPARS25 sampling sequence. Each spatial scan lasted 69.62 s, and the visit was preceded by a 1.71 s image collected in the F130N filter. All HST data used in this paper is archived at the Barbara A. Mikulski Archive for Space Telescopes (MAST): \dataset[https://doi.org/10.17909/fe8t-na27]{https://doi.org/10.17909/fe8t-na27}.

We used a version of the custom HST WFC3 data and light curve analysis pipeline described in \citet{Sheppard2021}, nicknamed \texttt{DEFLATE} (Data Extraction and Flexible Light curve Analysis for Transits and Eclipses)\footnote{The \texttt{DEFLATE} source code is available on Github at https://github.com/AstroSheppard/WFC3-analysis.}. \texttt{DEFLATE} uses the \textit{ima.fits} files from the MAST but separates the forward and reverse scans for independent processing, since the spatial scans tend to be offset in the spatial direction by several rows, complicating aperture determination. DEFLATE then eliminates the background noise in each exposure using the ``difference reads'' method \citep{deming2013}. While it is possible to use a scaled version of a master sky background file to remove specific background patterns \citep{gennaro2018}, the method we used takes advantage of the multiple readouts within each exposure to remove background in a purely data-defined way. As a final step, the pipeline propagates the uncertainty due to this background subtraction by adding it in quadrature, since the new count for each pixel is $F_{new} = F_{old} - F_{bkg}$. The difference-reads method lowers the likelihood of cosmic rays impacting the data (since the location of the source on the detector has no bearing on cosmic rays, any ray that hits a non-source pixel during the observation is automatically zeroed out). It also allows for resolving the source from companions or other field sources in the case of overlapping scans since the individual difference frames do not overlap.

Due to distortions, the pixel-to-wavelength calibration (i.e, wavelength solution) depends on the exact X and Y position on the detector, and so it varies between observations. Still, it is a roughly linear conversion that follows the following set of equations \citep{wakeford2013}:
\begin{equation}
    \lambda_{(X_{\textrm{ref}}, Y_{\textrm{ref}})} = \lambda_{\textrm{ref}} = a_0 + a_1*X_{\textrm{ref}}
    \label{eq:ref_wave}
\end{equation}
\begin{equation}
    \lambda_{\textrm{pixel}} = \lambda_{\textrm{ref}} + Y_{\textrm{dispersion}}*(X_{\textrm{pixel}} - X_{\textrm{shift}})
    \label{eq:dispersion}
\end{equation}

\begin{figure}
    \centering
    \includegraphics[width=0.5\textwidth]{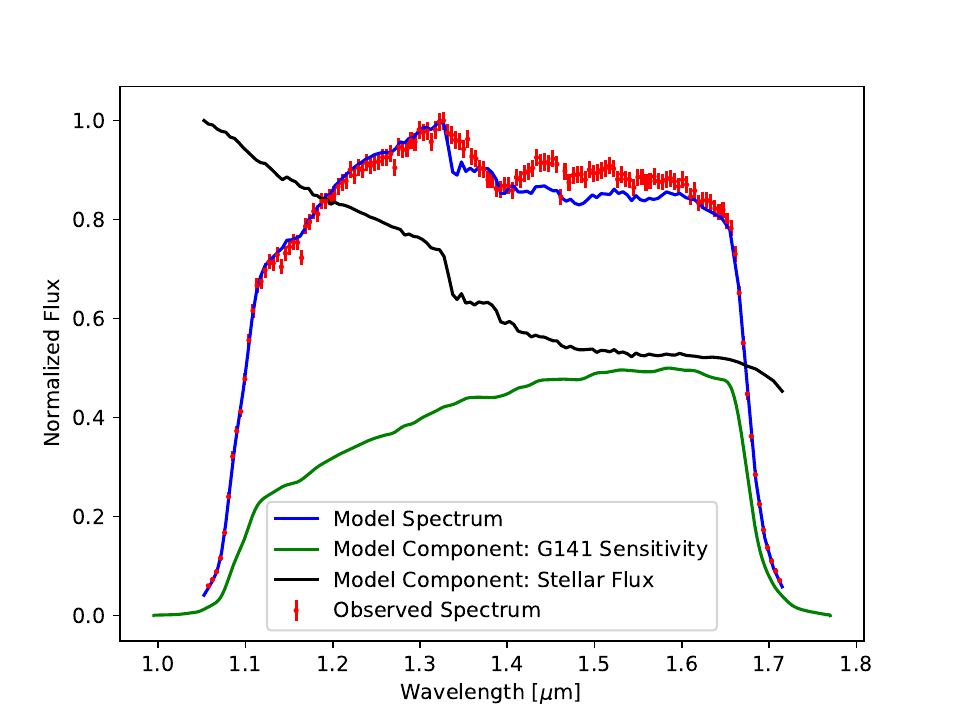}
    \caption{The model we generated for the stellar spectrum provides a sufficient match to the observed spectrum over the G141 wavelength range for use in our wavelength calibration}.
    \label{fig:wavefit}
\end{figure}

The reference coordinates $(X_{\textrm{ref}}, Y_{\textrm{ref}})$ were determined by the photometric images taken at the beginning of each visit. Coefficients for converting this reference pixel to a reference wavelength (a$_0$, a$_1$) were determined empirically by \citet[][Table 5]{kuntschner2009}. The wavelength of light recorded by a particular pixel is dictated by the dispersion for the Y-coordinate of the reference pixel (Y$_{\textrm{dispersion}}$) and the intrinsic offset (X$_{\textrm{shift}}$, in pixels) between the location of the filter image and the grism-dispersed light. Y$_{\textrm{dispersion}}$ and X$_{\textrm{shift}}$ are constrained, but spatial scan mode complicates those values. Consequently, \texttt{DEFLATE} fits for these values by comparing an observed out-of-transit spectrum to an ATLAS stellar model \citep{castelli2003} multiplied by the G141 grism sensitivity curve. The stellar model combines the line and continuum fluxes as ($\alpha\times$Line + Continuum), essentially allowing the strength of the stellar lines to vary to compensate for metallicity or other opacity mismatches. Figure~\ref{fig:wavefit} shows the result of the fit for L98-59~c.

% \begin{figure*}%[H]
%  {\caption[Example G141 Wavelength Calibrations]{\small Example wavelength calibration for three scenarios. \textbf{Top Left:} L9859 with fixed stellar model line-flux strength. \textbf{Top Right:} L9859 with scaled line-flux contribution. Though a better fit, the wavelength solution is nearly identical to the top-left case. \textbf{Bottom:} Same process (fixed line flux) for the simpler spectrum of the hotter WASP-79.}\label{fig:wavefit}}
%   {\includegraphics[width=0.48\textwidth, keepaspectratio]{DataAn_figures/l98_actual_wave_solution.pdf}
%   \includegraphics[width=0.48\textwidth, keepaspectratio]{DataAn_figures/l98_scaled.pdf} \\
%   \includegraphics[width=0.45\textwidth, keepaspectratio]{DataAn_figures/wasp79_wave_solution.png}}
% \end{figure*}

We determined the wavelength solution separately for both the forward-scan and reverse-scan light curves. The forward and reverse scans tend to be offset vertically from one another by a small amount, and the difference in wavelength solution is never more than 3\% and typically around 1.5\%, well within the size of a detector pixel (i.e, subpixel shift). We found that the wavelength solution was not significantly impacted by exact stellar model choice, error scaling, or line-strength scaling. 

\texttt{DEFLATE} uses the downloadable WFC3 flatfield files to divide out the flat-field from both the data and the error array (to propagate uncertainty), and removes the \texttt{cal-wfc3}-flagged ``bad'' pixels (which are identical across all exposures) by giving them zero-weight. It is possible to interpolate flux values at these pixels, but we preferred the zero-weight method since it requires fewer assumptions. The zero-weight pixels make up roughly 2\% of all pixels in an exposure. \texttt{DEFLATE} also uses a corrected median time filter to flag cosmic rays. Before applying the filter, \texttt{DEFLATE} normalizes each pixel by the median of its row, which prevents \texttt{DEFLATE} from flagging entire rows as cosmic rays since they are distorted by time-dependent instrumental effects, inconsistent spatial scan rates, and obviously the transit/eclipse itself. We then used a double-sigma cut, first applying an 8$\sigma$ cut to remove any extreme outliers, then applied a second 5$\sigma$ cut to correct the remaining energetic particles. Less than 0.5\% of all pixels in L~98-59~c's observations were impacted by cosmic rays, and typically only a few pixels per exposure were impacted.

Finally, \texttt{DEFLATE} follows a simple procedure to define a light curve extraction aperture. It first defines the maximum flux of an exposure as the median of the five rows with the greatest flux. The edge of the box is set to the outermost row and column with a median value of greater than 3\% of the maximum flux. This relatively low cut-off captures the entire first-order spectrum and minimizes the impact of vertical shifts. This method maximizes the SNR from the source and avoids over-processing the data. 

\section{Light Curve Analysis}
\label{sec:l98_lightcurve}

Modeling a transit light curve has two major components: modeling the physical transit, and modeling the non-astrophysical instrumental effects related to how the WFC3 instrument collects flux, i.e. the instrumental systematics. WFC3 observations commonly exhibit several instrumental effects; the most prominent effects are a hook/ramp feature due to charge-trapping, a visit-long decrease in flux, a ``breathing'' effect based on changing temperatures during HST's orbit, and a wavelength jitter effect \citep[e.g,][among many others]{Berta2012,Wakeford2016,zhou2017,Tsiaras2018}. These features vary in magnitude across different observations in non-obvious ways. There is no encapsulating physically-motivated model to describe all of these effects (though recently individual features have been modeled more successfully, e.g. \citet{zhou2017}). Instead of using inherent properties of the detector, these features are typically removed using empirical methods \citep{gibson, nikolov2014, haynes2015}. For this work, we used a new version of parametric marginalization, a Bayesian model averaging strategy that was conceptually introduced to exoplanet light curves by \citet{gibson2014} and first applied to WFC3 transit spectroscopy by \citet{Wakeford2016}.

Similar to \citet{Wakeford2016}, we found that fourth-order polynomial parameterization consistently described the systematic effects while preserving computational time. We used a grid of models that included up to four powers of phase ($\phi_{\textrm{HST}}$) and four orders of wavelength shift. Of the 5 forms of orbital phase-dependent visit-long slopes (none, linear, quadratic, exponential, and log), we only allowed linear, as discussed in the next section. Each higher power included all lower powers (e.g, 3rd order $\phi_{\textrm{HST}}$ phase is a$_0\times$$\phi_{\textrm{HST}}$ + a$_1\times$$\phi_{\textrm{HST}}^{2}$ + a$_2\times$$\phi_{\textrm{HST}}^{3}$), and there were no cross terms. This resulted in a grid of 25 systematic models (5 possible $\phi_{\textrm{HST}}$ powers $\times$ 5 possible shift powers $\times$ 1 possible slope parameterization). There were two additional parameters: separate normalization constants for the forward (A$_f$) and reverse scans (A$_r$). 
%Removed, exact text from Sheppard: It is typical for the two directions to be offset, though that is the primary effect and they can still be fit simultaneously.

To efficiently obtain light curve parameter values and uncertainties, we followed the methodology presented in \citet{Sheppard2021} and fit each model using \texttt{KMPFIT}\footnote{\texttt{KMPFIT} is available as part of the Kapteyn Package available at \url{https://github.com/kapteyn-astro/kapteyn/}}.
Each model was then weighted by its Bayesian evidence and marginalized over the model grid.

\begin{figure}
\centering
  \includegraphics[width=0.5\textwidth]{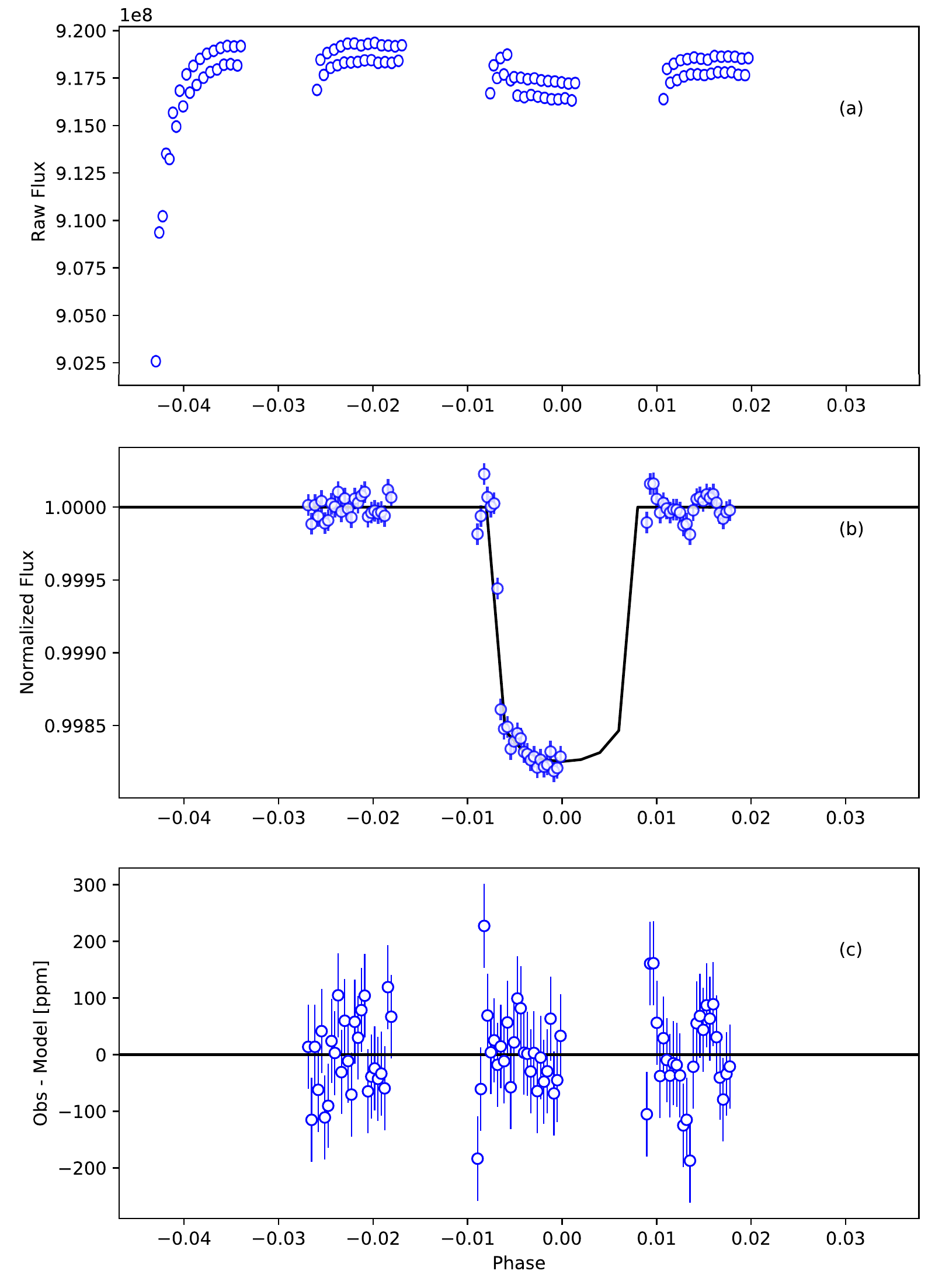}
  \caption{\small Visualization of white light curve fit for the highest weighted systematic model for L~98-59~c demonstrates a good model fit to the observed data. Panel (a) shows the band-integrated light curve. Panel (b) shows the detrended light curve as well as the best-fitting transit model. The instrumental effects and transit model parameters were fit for simultaneously. Panel (c) shows the residuals between the data and the best-fitting model.}\label{fig:l98_white_curves}
\end{figure}

% \begin{figure}
% \centering
% \includegraphics[width=.45\textwidth,]{DataAn_figures/marg_corner_astrophysical.pdf}
% \caption{Corner plot of astrophysical parameters for an MCMC fit of the highest weighted systematic model. The model converged and we derived uncertainties slightly less than \textbf{those from KMPFIT; however, the overall white light uncertainty in Table~\ref{tab:l98_derived} is larger due to the impact of multiple models in the marginalization. No correlation was seen with any non-astrophysical lightcurve parameters.}}\label{fig:l98ccorner}
% \end{figure} 

We used the \texttt{BATMAN} software \citep{kreidberg2015a} to generate transit models that were used for the astrophysical component of the light curve model. The model strongly constrained the transit depth (R$_p$/R$_s$) and center of transit time (T$_0$) with weak constraints on scaled semi-major axis (a/R$_s$), inclination (i), and a linear limb-darkening coefficient (c$_0$). We assumed a four-parameter nonlinear limb darkening (LD) and derived the coefficients by interpolating values from \citet{claret2012} to the central wavelength of WFC3 (1.4\,$\mu$m). These coefficients were fixed for light curve fitting as HST's poor phase coverage does not well constrain the shape of transit. We also compared the LD values with earlier models by \citet{claret2011} to be sure results are not sensitive to LD source and also tested a linear LD law with the coefficient being a fitted parameter.

We first fit a model to the white light curves, which provided a confidence check on the data, maximized SNR for deriving wavelength-independent properties such as inclination and a/R$_s$, and captured the structure of residuals for each systematic model, if present. Determining the residuals allowed further de-trending of spectral curves via white light residual removal \citep{mandell2013,deming2013,Haynes:2015cf}. While this method is suitable for M-dwarf targets that are near constant flux across the bandpass (including L98-59, despite the small variations in flux we see in Figure~\ref{fig:wavefit}), this method is not generally applicable to hotter stars. We fixed the orbital period, inclination, and $a/R_s$ to their values in Table~\ref{tab:l98_lit}, and only allowed linear visit-long slopes, since the L~98-59 dataset only had three usable orbits covering a small amount of the out-of-transit baseline. 

% Exact text from Sheppard: As is common practice, we ignore the systematic-dominated first orbit in the white light analysis; however, the use of common-mode detrending provides the option of including that orbit in the spectral light curve analysis. 
The unprocessed light curve, along with the light curve with instrumental systematics removed, and the residuals when the highest-weight systematic model was subtracted from the instrumentals-subtracted light curve are shown in Figure~\ref{fig:l98_white_curves}. The derived transit depths and center-of-transit times (T$_c$) are given in Table~\ref{tab:l98_derived}. The white light depth was measured to be $1620\pm{24}$~ppm. The derived depths are insensitive to model assumptions, varying less than 10~ppm when either a linear LD or $a/R_s$ was included in the model as fitted parameters (significantly less than 1$\sigma$), or if a quadratic visit-long slope was assumed. The reduced chi-squared of each fit was around 1.2, which is typical of HST white light curves.

\begin{table}
\centering       
\caption{L~98-59~c White Light Curve-Measured Transit Parameters}
\label{tab:l98_derived}
\begin{tabular}{l c c}   
\hline    
    Observation & Transit Depth & T$_c$ \\
     & [ppm, $(R_p/R_s)^2$] & [BJD-2457000] \\\hline
    L~98-59~c & $1620\pm{24}$  & $1946.7068 \pm{0.0001}$ \\
\hline
\end{tabular}
\end{table}

%We also show the corner plot \citep{foremanmackey2016} of astrophysical parameters in Figure~\ref{fig:l98ccorner}. 
%
% The model converged and we find the posteriors of the model, such as transit depth, are Gaussian.

% and we derived uncertainties slightly less than those from KMPFIT (shown in Table~\ref{tab:l98_derived}. The two methods are in excellent agreement, down to the ppm: both find a depth of 1620ppm and the MCMC uncertainty is 98\% of the KMPFIT uncertainty, with a posterior Gaussian distribution. However, when the impact of using multiple models in the KMPFIT method is included, that inflates the uncertainty from 11 ppm to 24 ppm. We have adopted this larger uncertainty.

% ; however, the overall white light uncertainty in Table 2 is larger due to the impact of multiple models in the marginalization. No correlation was
% seen with any non-astrophysical lightcurve parameters
% 
To further validate these results, and to make sure the derived uncertainties are reasonable, we fit the highest-weighted systematic model of L~98-59~c using Markov-chain Monte Carlo \citep[\texttt{emcee};][]{foremanmackey2013}. The model successfully converged, and we found that the posteriors, such as the transit depth, exhibit Gaussian distributions. We derived uncertainties slightly smaller than those obtained with KMPFIT, which are shown in Table~\ref{tab:l98_derived}. The two methods are in excellent agreement, down to the parts-per-million (ppm) level: both estimate a depth of 1620 ppm, with the MCMC uncertainty being 98\% of the KMPFIT uncertainty. However, when the impact of using multiple models in the KMPFIT method is included, the uncertainty inflates from 11 ppm to 24 ppm. We have adopted this larger uncertainty. Notably, the overall white light uncertainty in Table~\ref{tab:l98_derived} is larger due to the effects of model marginalization. Correlations were not seen with any non-astrophysical light curve parameters.

\begin{figure}%[p]
\centering
{
\includegraphics[width=.5\textwidth,]{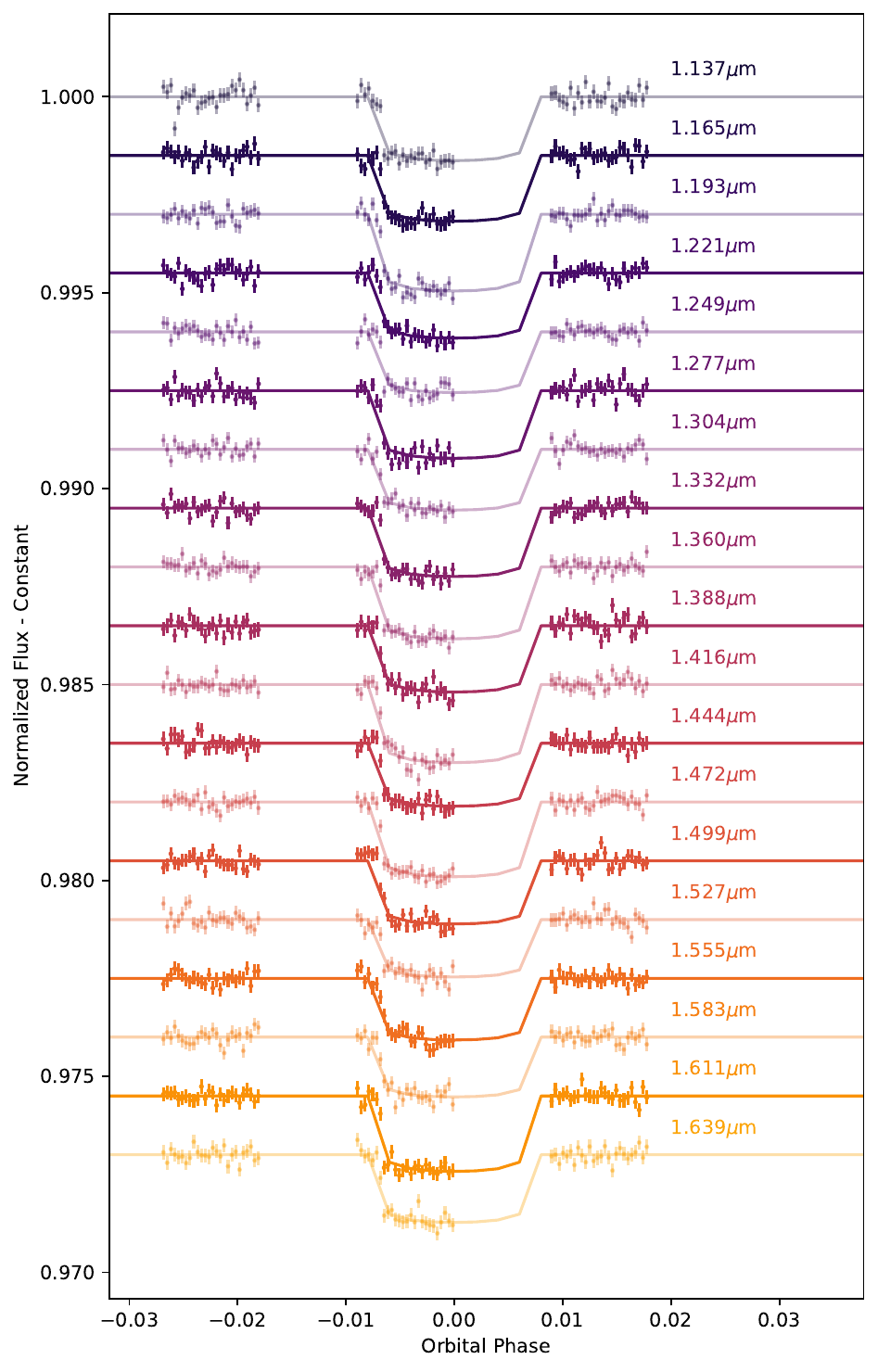}
\caption{Spectral light curves for L~98-59~c, collected during the first visit; systematic variations have been removed.}
\label{fig:l98c_spectral}
}
\end{figure} 

% \subsection{Transit Spectra Derivation}

\begin{table}[t]

\centering       
\caption{Transmission Spectra of L~98-59~c, with bin size 0.0279 $\mu$m and resolution $\sim50$.}
\label{tab:l98_spec}

\begin{tabular}{l c}     % 4 columns 
\hline      
    $\lambda$ [$\mu$m] & Depth [ppm] \\
\hline
    1.123--1.151 & 1561 $\pm$ 56  \\  
    1.151--1.179 & 1609 $\pm$ 55  \\
    1.179--1.207 & 1686 $\pm$ 54 \\
    1.207--1.235 & 1600 $\pm$ 52  \\
    1.235--1.263 & 1616 $\pm$ 51 \\ 
    1.263--1.291 & 1539 $\pm$ 59  \\
    1.291--1.318 & 1585 $\pm$ 49  \\
    1.318--1.346 & 1572 $\pm$ 51  \\
    1.346--1.374 & 1658 $\pm$ 53  \\
    1.374--1.402 & 1628 $\pm$ 56 \\
    1.402--1.430 & 1693 $\pm$ 56  \\
    1.430--1.458 & 1697 $\pm$ 55  \\
    1.458--1.485 & 1721 $\pm$ 53  \\ 
    1.485--1.513 & 1665 $\pm$ 52 \\
    1.513--1.541 & 1549 $\pm$ 54  \\
    1.541--1.569  & 1662 $\pm$ 54  \\ 
    1.569--1.597 & 1625 $\pm$ 54 \\
    1.597--1.625 & 1739 $\pm$ 55  \\
    1.625--1.653 & 1635 $\pm$ 54  \\ 
\hline
\end{tabular}
\end{table}

\begin{figure*}
\centering
\includegraphics[width=0.95\textwidth]{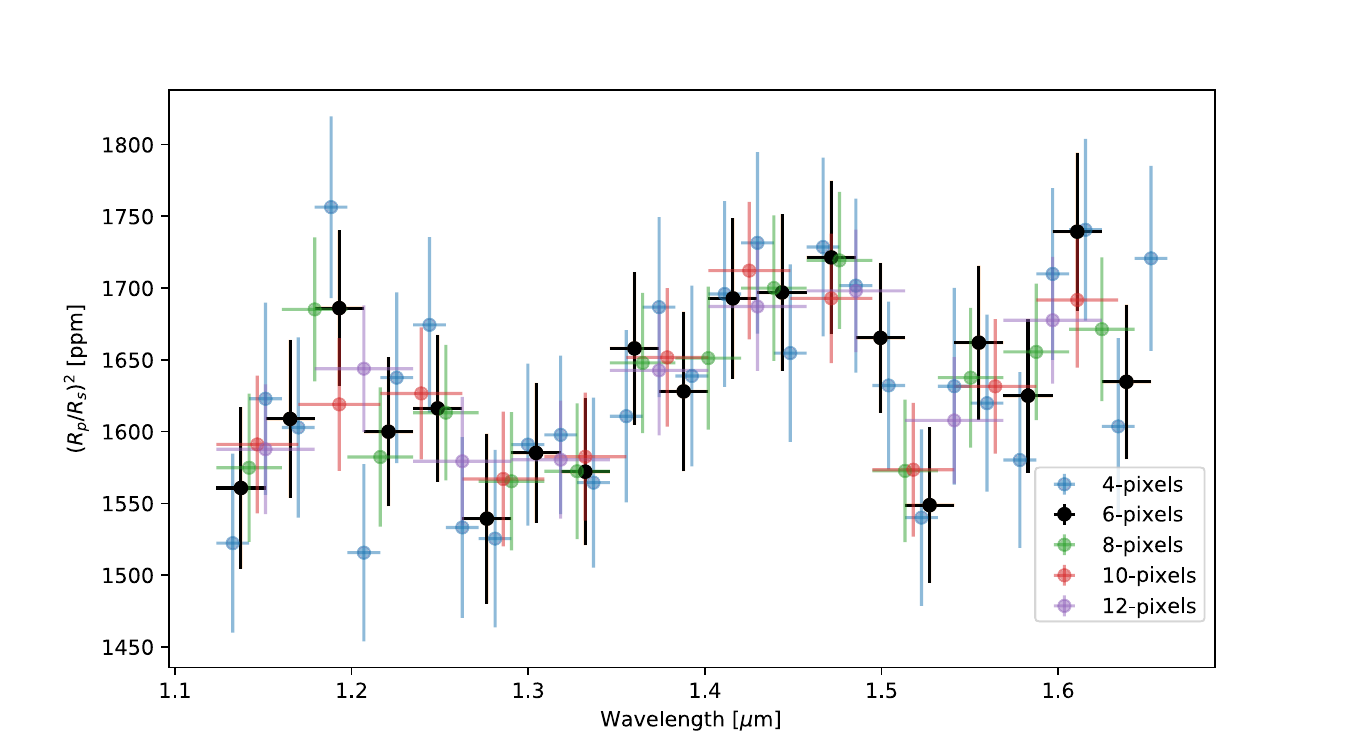}
\caption{The marginalization-derived transit spectrum for L~98-59~c at different resolutions all show evidence for a non-flat spectrum. The shape of the spectrum is not sensitive to the spectral bin size.}
\label{fig:l98c_spectrum}
\end{figure*}

We binned the 1-D spectra from each exposure in the region covered by the grism response curve (1.1--1.6\,$\mu$m) to compute spectral-photometric time-series for each spectral bin. We tested several bin widths since the long scan observations (close to 300 rows) are more at risk of wavelength blending \citep{tsiaras2016}, which will effect larger bins less than smaller ones. We find no difference as a function of bin width (see Figure~\ref{fig:l98c_spectrum}), and we choose 6-pixel-wide (0.0279 $\mu$m) bins to maximize resolution without drowning the signal in noise. The spectrum is given in Table~\ref{tab:l98_spec}, and the spectral light curves are shown in Figure~\ref{fig:l98c_spectral}.

% \subsubsection{WFC3 Transit Spectrum Verification}
% \label{subsec:verify}

Following \citep{Sheppard2021} we looked into three different methods to assess the robustness of our analysis: (1) the goodness-of-fit of the highest-weighted systematic model for each light-curve using the reduced $\chi^2$ statistics, (2) a residual normality test, and (3) whether red noise was present in the light curve residuals, as that could bias inferred transit depths \citep{cubillos2017}.

\begin{figure*}%[p]
\centering
\includegraphics[width=\textwidth]{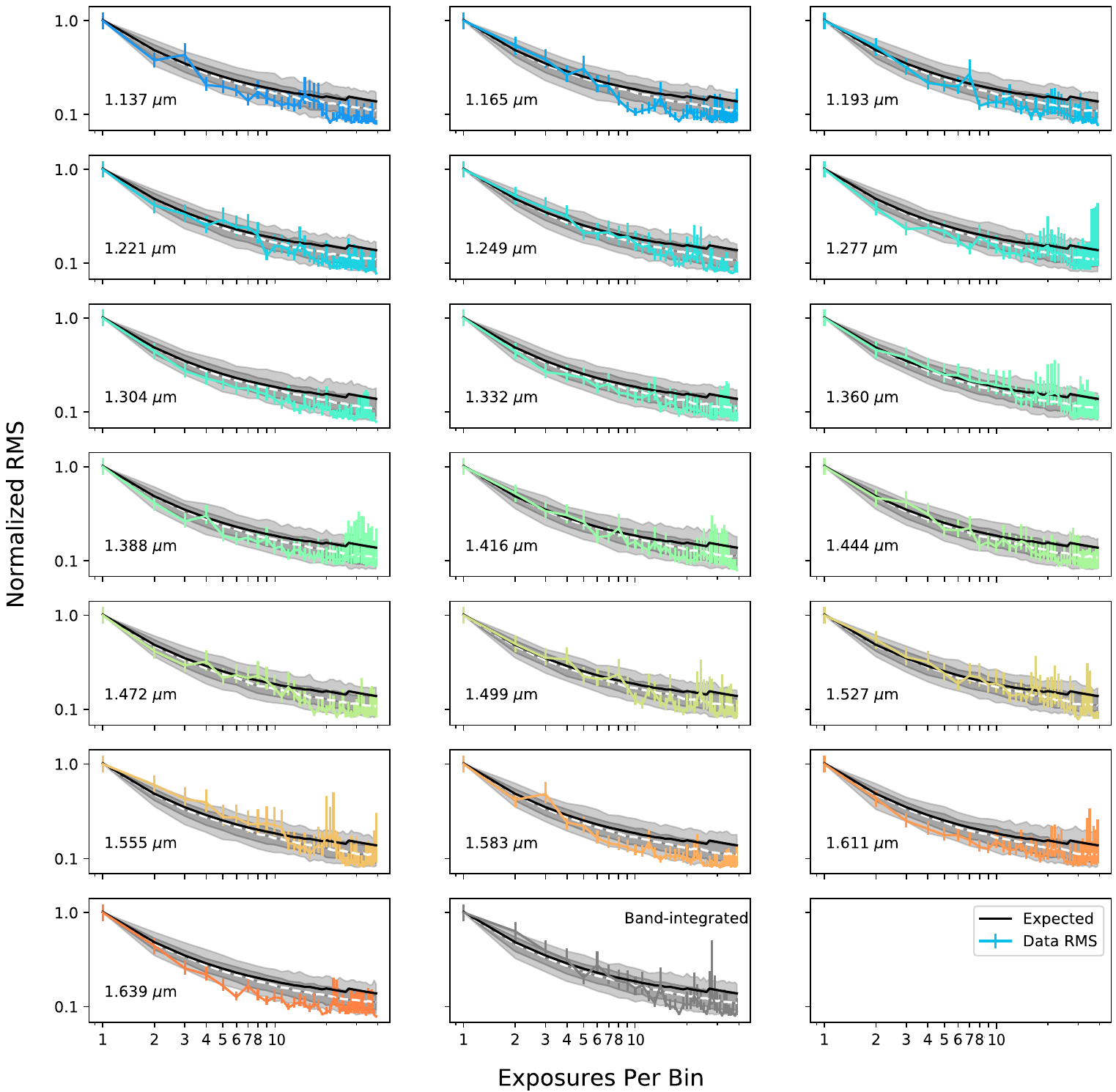}
\caption{Binned RMS analysis for each spectral bin showed no evidence of red noise. The RMS of the data residuals are shown by the colored lines. The solid black line is the theoretical trend from \citet{cubillos2017}. The dashed-white line is the median value from simulated pure white noise residuals. The gray regions are the 1- and 2-$\sigma$ ranges for the simulated white noise residuals, which demonstrates that the observed data is consistent with largely uncorrelated (e.g. white noise).}
\label{fig:l98c_rednoisea}
\end{figure*} 

\begin{figure*}%[p]
\centering
\includegraphics[width=\textwidth]{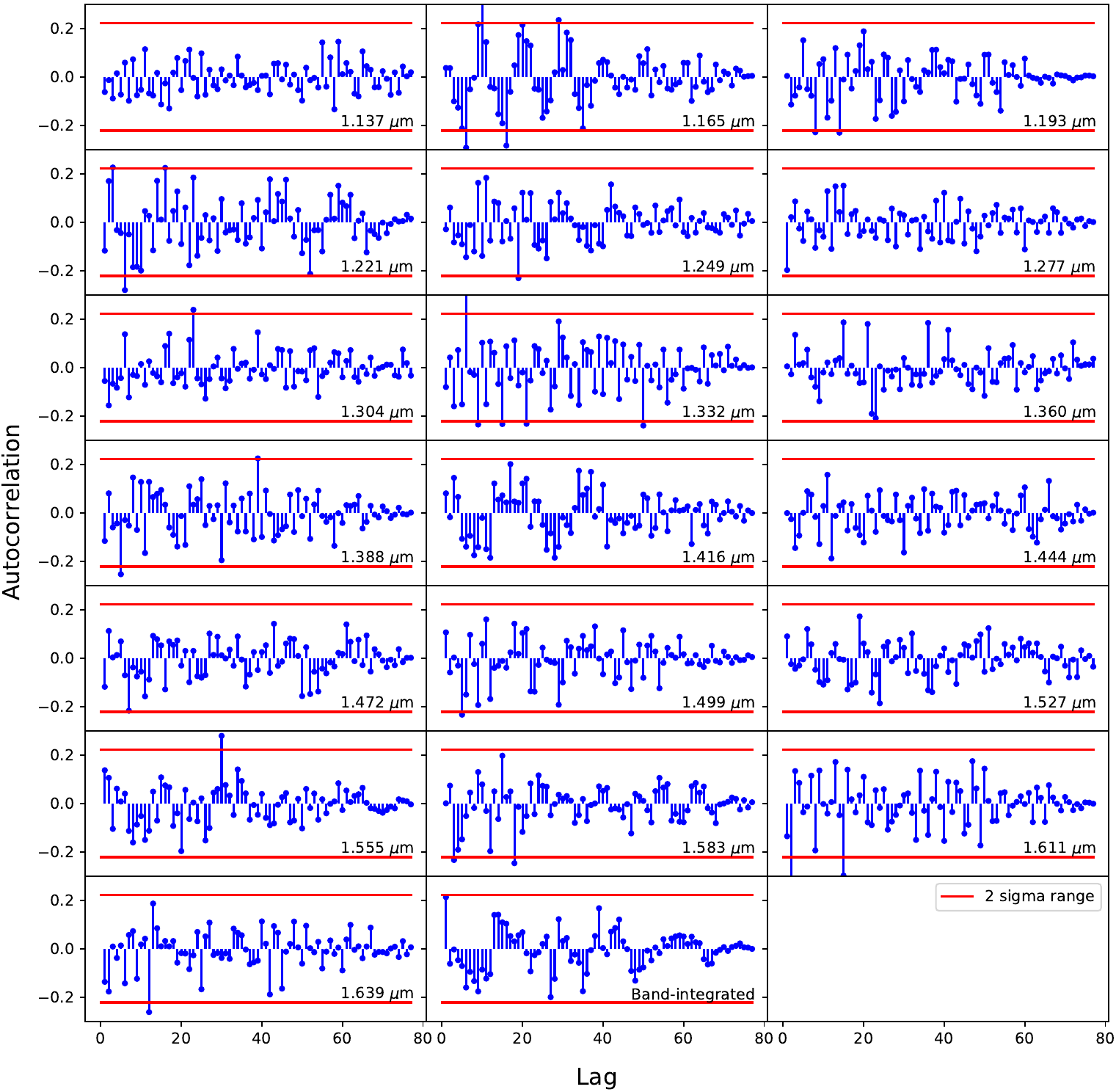}
\caption{The autocorrelation function of the residuals for each spectral bin.} The blue lines and dots show the autocorrelation function as a function of lag, with lag 0 left out for clarity. The solid red lines indicate the 2$\sigma$ range: autocorrelation value within these lines are not considered significant. The observed data has minimal significant autocorrelation.
\label{fig:l98c_rednoiseb}
\end{figure*} 

For both the band-integrated (white light) and spectrally binned light curves  we measured reasonable reduced $\chi^2$ values that ranged between 0.7 and 1.4.
The band-integrated analysis ($\chi^2_{\nu}=1.2$) and all spectral bins (median $\chi^2_{\nu}=0.9$) fall within the expected range, except for the 1.499\,$\mu$m light curve. The reduced $\chi^2$ of this bin is 0.59 which likely indicates that the uncertainties in this light curve are overestimated, probably due to incorporating white light residuals that inflated uncertainties.

A residual normality test assess whether the residuals for a model are Gaussian-distributed. We used the \texttt{scipy} implementation of the Shapiro-Wilk test for normality \citep{shapiro65, scipy}. Normality is rejected at the $5\%$ significance level only for the 1.14\,$\mu$m spectral bin residuals, and in this case this is owing to a single outlier. Removing this outlier enabled us to recover a consistent depth and uncertainty and the residuals are consistent with normality. Further, ignoring residuals recovered almost the exact same transit depth without any normality issues. We retained this exposure in our analysis.

Finally, we tested for correlated noise in the residuals following the time-average methodology of \citet{cubillos2017} (also see \citet{pont2006}) and using MC$^3$\footnote{MC$^3$ software is available from \url{https://github.com/pcubillos/mc3}
}. Noise can be thought of as the sum of a purely random (white) noise and a time-correlated (red) noise: $\sigma_{total} = \sqrt(\sigma^2_w/N + \sigma^2_r)$ \citep{pont2006}. As normally distributed residuals with mean of zero (i.e, if uncorrelated white noise is the dominant uncertainty source) are averaged in time, the scatter in the points decreases proportional to $\sigma_{w}/\sqrt{N}$. If red noise is significant, then the time averaging only decreases noise until it flattens out at $\sigma_r$. One can test for the impact of red noise by time-averaging the residuals and comparing the resulting RMS function to theoretical expectations of white noise. 
%For example, first average each point with its neighbor, then recalculate the RMS of those averaged points. 
We enhanced this method by simulating synthetic normally-distributed residuals with the same standard deviation as the actual residuals, and putting them through the same method (Figures~\ref{fig:l98c_rednoisea} and ~\ref{fig:l98c_rednoiseb}). We note the 1 and 2$\sigma$ bands for random, pure white noise residuals compared to the results for the actual residuals. For every bin, the residuals are consistent with random white noise for every bin size. We find no evidence of correlated noise.

Figure~\ref{fig:l98c_rednoiseb} shows a visualization of the correlated noise test by looking at the autocorrelation function of the residuals. This method is not purely quantitative, but can provide another look at potential structure in the residuals. The red lines in Figure~\ref{fig:l98c_rednoiseb} indicate the 2$\sigma$ line -- roughly indicating ``significant'' correlations at that lag. A few lags crossing this threshold is not concerning, since 2-sigma events happen roughly 5\% of the time and we are sampling many bins. This is less quantitative, but autocorrelation functions that appear too structured can be, unsurprisingly, indicative of structured noise. An example is the 1.165 $\mu$m bin which resembles a decreasing sinusoid; however, structure below significance is less problematic. Different systematic model selections from the highest-ranking models give the same transit depth at this bin without structured residuals, and it passes the other red noise test, so we have confidence in the depth determination.

% exact text as Sheppard: With the caveats noted above, marginalization does an excellent job in fitting the spectral light curves. Together, these tests support the validity of the derived transit depths and uncertainties.

\section{Exploratory Analysis of Potential Atmospheres with PLATON}
\label{sec:l98_results}

In the subsequent sections, we study whether the apparent structures in the spectra of L~98-59~c are significant and indicative of an atmosphere with a reasonable chemical composition. Only a low mean-molecular weight atmosphere could produce spectral features with amplitudes similar to the features seen in our results (roughly 100 ppm). However, due to the relatively large error bars on each individual spectral bin, it is not obvious that a model with molecular features would be a significantly improved fit over a straight line (indicative of no atmosphere, an atmosphere with a high mean molecular weight, or high-altitude clouds). %Further, stellar activity can potentially contaminate transit spectra and mimic molecular features \citep{Barclay2021}.

We utilized two different strategies to investigate the potential for detecting an atmosphere. We first used the open-source retrieval tool PLATON \citep{zhang2019} to perform a Bayesian statistical retrieval of the atmospheric parameters assuming a H$_{2}$-rich composition (about 1--2\% by mass \citep{Lopez2014}) and equilibrium chemistry. We then examined the ability to constrain the presence of individual molecular constituents using a more simplistic fitting scheme for combinations of individual absorbers with the Planetary Spectrum Generator \citep[PSG,][]{PSG}.

\subsection{PLATON Atmospheric Modeling and Retrieval Methodology}
\label{sec:l98_plat}

PLATON is open-source retrieval software developed by \citet{zhang2019}, which comprises a forward model and an algorithm for Bayesian inference. Though there are minor differences, it essentially uses the same forward model as Exo-Transmit \citep{kempton2017}. The software assumes an H$_2$-He/dominated atmosphere, and though it has recently incorporated free retrieval capabilities, in this study we exclusively used the version which assumes chemical equilibrium. Though these constraints naturally limit the types of planetary atmospheres that can be explored, it was useful in contextualizing the spectrum and investigating the likelihood of an H$_2$-dominated atmosphere on L~98-59~c.

Table~\ref{tab:l98_priors} describes the parameters and their priors for the PLATON atmospheric retrieval. We allowed planet radius, C/O, metallicity, temperature, and cloudtop pressure to vary, and assumed an isothermal temperature profile. C/O and metallicity dictate the elemental ratios in the atmosphere, which are input with temperature into a chemical equilibrium code \citep[ggchem;][]{woitke2018} to determine the abundance of every species at every pressure layer. Each chemical parameter was given a prior set by computational limits (most notably T$_{min}$=300~K), and the mass/radius priors were set by literature values \citep{Demangeon2021}; we included stellar radius and planetary mass in order to propagate the literature uncertainties forward. The retrieval utilized nested sampling \citep{skilling2004, speagle} with 200 live points to sample the parameter space and calculate a Bayesian evidence for the model. %The assumed C/O ratio is the Solar value.

% \begin{table*}
% \centering       
% \caption{Priors for parameters used in L~98-59~c Retrievals}
% \label{tab:l98_priors}

% \begin{tabular}{l l l r}     % 4 columns 
% \hline   
%     Parameter & Symbol & Prior Distribution\\
% \hline
%     Planet Radius & $R_p$ & $\mathcal{U}(0.68, 2.03)$ & 1.35~$R_{\textrm{Earth}}$ \\ 
%     Limb Temperature & $T$ & $\mathcal{U}(300, 1100)$& 550~K\\  
%     Carbon-oxygen ratio & C/O & $\mathcal{U}(0.05, 2.0)$ & 0.53\\ 
%     Metallicity & $Z$ & $\mathcal{LU}(-1, 3)$ & 1 $Z_{\odot}$\\  
%     Planet Mass & $M_p$ & $\mathcal{N}(2.40, 0.35)$ & 2.4~$M_{\textrm{Earth}}$ \\
%     Stellar Radius & $R_s$ & $\mathcal{N}(0.314, 0.01)$ & 0.314~$R_{\odot}$ \\
%     Cloudtop Pressure & $P_{\textrm{cloud}}$ & $\mathcal{LU}(-3, 8)$ & 1~Pa\\ 
%     Stellar Effective Temperature & $T_{\textrm{star}}$ & Fixed & 3429~K\\
%     Spot Temperature & $T_{\textrm{spot}}$ & Fixed& 2920~K \\
%     Spot covering fraction & $f_{\textrm{spot}}$& $\mathcal{U}(0, 0.5)$ & 0.1\\
   
% \hline
% \end{tabular}
% \end{table*}

\begin{table*}
\centering       
\caption{Priors for Parameters Used in L~98-59~c Equilibrium Retrievals}
\label{tab:l98_priors}
\footnotesize
\begin{tabular}{l l l}     % 4 columns 
\hline   
    Parameter & Symbol & Prior Dist.\\
\hline
    Planet Radius  [$R_{\oplus}$]& $R_p$ & $\mathcal{U}(0.7, 2.1)$ \\ 
    Limb Temperature [K]& $T$ & $\mathcal{U}(300, 1100)$\\  
    Carbon-oxygen ratio & C/O & $\mathcal{U}(0.05, 2.0)$\\ 
    Metallicity & $Z$ & $\mathcal{LU}(-1, 3)$\\  
    Planet Mass [$M_{\oplus}$]& $M_p$ & $\mathcal{N}(2.22, 0.26)$\\
    Stellar Radius [$R_{\odot}$]& $R_s$ & $\mathcal{N}(0.30, 0.02)$\\
    Cloudtop Pressure [Pa]& $P_{\textrm{cloud}}$ & $\mathcal{LU}(-3, 8)$\\ 
    Stellar Effective Temperature [K]& $T_{\textrm{star}}$ & Fixed (3429~K)\\
    Spot Temperature [K]& $T_{\textrm{spot}}$ & Fixed (2920~K)\\
    Spot covering fraction & $f_{\textrm{spot}}$& $\mathcal{U}(0, 0.5)$\\
\hline
\footnotesize{$\mathcal{U}$ refers to a Uniform prior, $\mathcal{LU}$ a Log Uniform prior, and $\mathcal{N}$ a Normal prior.}
\end{tabular}
\end{table*}

\subsection{PLATON Retrieval Results}
\label{sec:l98_plat_results}

\begin{figure*}%[p]
  \includegraphics[width=\textwidth]{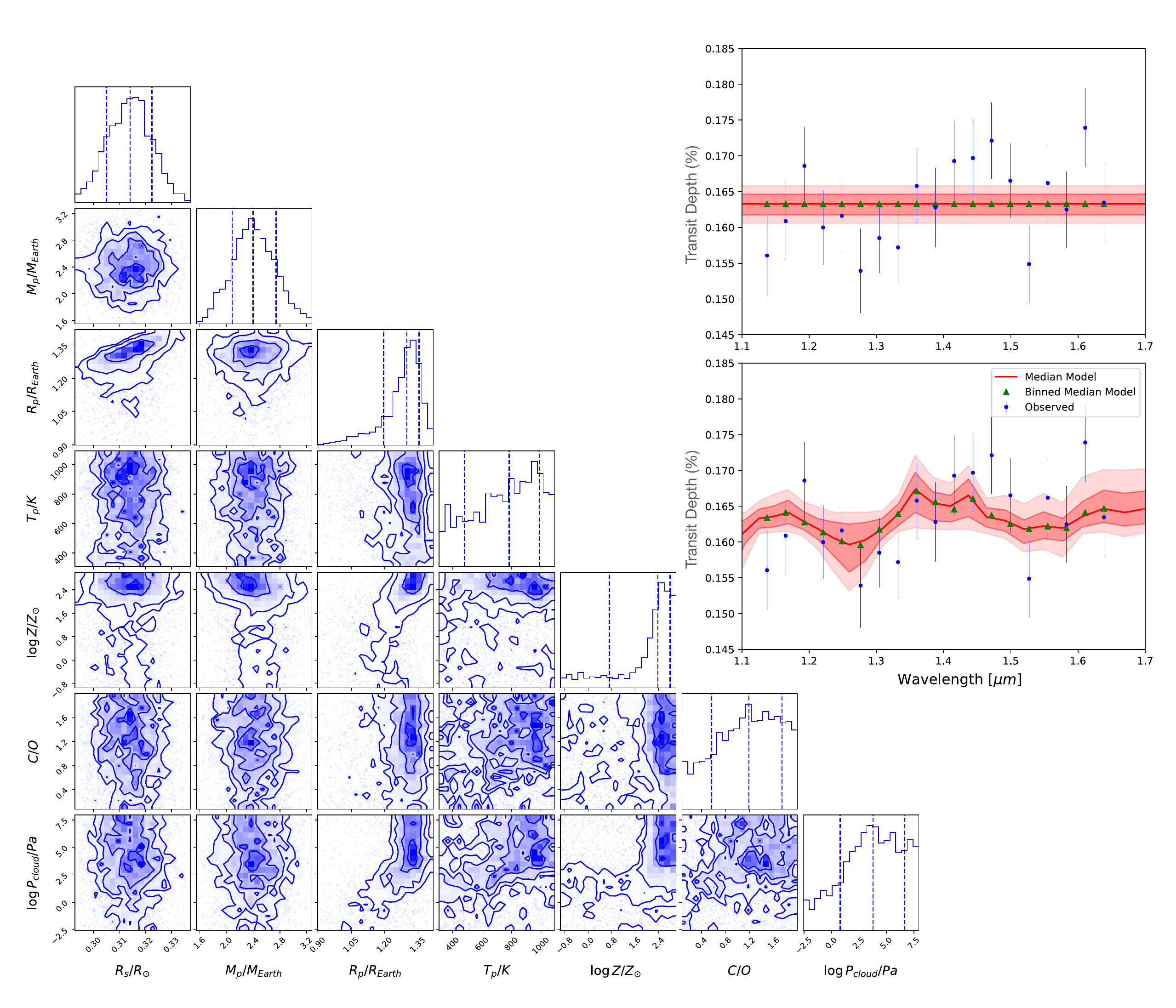}
  \caption{Retrieval results for L~98-59~c. \textbf{Left:} Corner plot for the best-fit fiducial H$_2$-dominated atmosphere. \textbf{Upper right:} Best fit assuming a flat spectrum (high clouds or no detectable atmospheric features). \textbf{Lower right:} Model spectrum assuming the median retrieved parameter values (green triangles) with 1 and 2-sigma contours (red) plotted over data (blue). }\label{fig:l98_retrieve}
\end{figure*}

The retrieval finds a best model fit with $\chi^2_{\textrm{Red}}$ = 1.15, with eight degrees of freedom. The resulting posterior distributions and best-fit model spectra from the retrieval are shown in Figure~\ref{fig:l98_retrieve}. Under the assumption that L~98-59~c has a H$_2$-dominated atmosphere with no disequilibrium processes, L~98-59~c was best described as a high-metallicity atmosphere (Z$\sim250\times$Z$_\odot$) with a likely super-solar C/O ratio. The retrieved atmospheric metallicity was consistent with predictions from the hypothesized mass-metallicity relationship from the Solar System planets \citep[e.g,][]{fortney2013}, and the retrieved R$_p$ was consistent with the literature (1.30$\pm$0.07~R$_\oplus$). The atmospheric temperature was poorly constrained, which is expected for a retrieval of transmission spectrum, and the cloud-top pressure was also weakly constrained due to the narrow wavelength coverage and the large uncertainties on the data. The best-fit model yielded small water features at 1.4 $\mu$m and 1.1 $\mu$m, but there was no statistically significant water detection. We note that the inferred water feature at 1.4 $\mu$m is roughly 100 ppm, which is consistent with predictions of the feature size for a H$_2$-dominated atmosphere, assuming 4 scale heights (derived from \citet{kreidberg2018}), but somewhat larger than what we would expect at 2 scale heights \citep{fu2017,Iyer2016,Wakeford2019,Gao2020}.

PLATON also allows for model comparison, since nested sampling naturally calculates the Bayesian evidence of a model. Although this evidence cannot act as an absolute goodness-of-fit metric, the ratio of two evidences provides a straightforward measure of how much more likely one model is in comparison to the other. This ratio is known as the odds ratio ($\mathcal{O}_{12}=\mathcal{Z}_1/\mathcal{Z}_2$), and is directly interpreted as ``Model 1 is $\mathcal{O}_{12}\times$ more probable than Model 2''. There are also empirically-determined benchmarks for converting $\mathcal{O}$ into more familiar $\sigma$-level significance \citep{trotta2008, Benneke2013}.

To determine the likelihood of an H$_2$ atmosphere on L~98-59~c, we compared the evidence of the retrieved fiducial atmosphere to that of a flat line spectrum. We introduced a flat spectrum into PLATON by fixing a very high, grey cloud. We then only allowed planet radius to vary in the fit. The resulting fit is shown in the upper-right panel of Figure~\ref{fig:l98_retrieve}. The odds ratio between the fiducial model and the flat-spectrum model was 3, which corresponds to a ``weak'' detection of roughly 2.1$\sigma$ in favor of the fiducial model over the flat-spectrum model, or about 75\% probability. We note that the specific sigma significance is relatively imprecise, since even a small numerical error in $\mathcal{Z}$ -- which is common \citep{speagle} -- could shift the odds ratio slightly below the empirical cut-off. We also compare the Bayesian evidences between the fiducial atmosphere and the same atmospheric model with no water opacity, finding the odds ratio to be $\sim$1. This indicates that there is no conclusive evidence of the presence of water vapor in the atmosphere. The evidence of water vapor specifically is weaker than that of the full atmosphere model since other opacity sources (e.g, NH$_3$) can ``fill in'' for water vapor and capture some of the structure in the observed spectrum.

\section{PSG Atmospheric Modeling and Retrieval}

The PLATON forward model and retrieval framework includes a number of assumptions about the atmospheric composition and structure that limit the ability to examine the evidence for individual atmospheric absorbers. For these reasons we also opted for an exploratory model-fitting approach using the Planetary Spectrum Generator (PSG), in which we analyze the potential for a molecular detection by looking at how well simulated spectra with a number of potential atmospheric absorbers fit the data. To quantify the goodness of the fit we used the reduced $\chi^2$ ($\chi^{2}_{red}$) of the data and the model, and we examined the $\chi^{2}_{red}$ across a range of molecular combinations and abundances.

PSG \citep{PSG,PSGbook} is a radiative transfer model and tool for synthesizing/retrieving planetary spectra (atmospheres and surfaces) for a broad range of wavelengths (50 nm to 100 mm, UV/Vis/near-IR/IR/far-IR/THz/sub-mm/Radio) and includes pre-defined templates for a large variety of instruments and observatories. As part of the retrieval framework of PSG, the tool has access to Nested Sampling and Optimal Estimation retrieval algorithms \citep{Rodgers2000} to analyze planetary data and retrieve atmospheric/surface/physical parameters of interest via minimization of spectral residuals.

Simulated spectra with PSG include molecular, atomic, aerosol and continuum (e.g., Rayleigh, Raman, CIAs) radiative and scattering processes, which are implemented via a layer-by-layer framework. Many spectral databases are available in PSG, but for this study we have employed the molecular parameters from the latest HITRAN-2020 database \citep{hitran20} that are implemented using a correlated-k method (21 molecular species are included). A list of the molecular and atomic databases (including HITRAN) are provided at \url{psg.gsfc.nasa.gov}, along with the specific molecules used, assumptions made, capabilities, and references. The HITRAN molecular database is complemented in the UV/optical with cross-sections from the Max Planck Institute of Chemistry database \citep{Kellerrudek13}. Besides the collision-induced absorption (CIA) bands available in the HITRAN database, the MT\_CKD water continuum is characterized as H$_2$O–H$_2$O and H$_2$O–N$_2$ CIAs \citep{kofmanvilla21}. %The PSG also includes a large database of extinction properties from many known aerosols, both measured in situ on Earth and laboratory-measured values for other planetary components \citep[see details in][]{PSG}. 

For this simulation we omitted aerosols, (clouds, haze, etc) and removed all molecules other than the two we wished to explore for any given simulation. The structure of the atmosphere was described in PSG by specifying for each layer the pressure (bars), temperature (K), and the abundances of atmospheric constituents with respect to the total gas content. For each gas, layer-by-layer integrated column densities (molecules m$^{-2}$) were then computed along the transit slant paths employing a pseudo-spherical and refractive geometry. We assumed the molecule abundance was held to a consistent volume mixing ratio throughout the atmosphere; however, the temperature-pressure profile was defined using the \cite{Parmentier2014} analytic grey opacity model.

\subsection{Chi-Squared Analysis Methodology}
As previously pointed out, the large error affecting each spectral bin may be too large to clearly discern the abundance of gaseous species of interest by working with a classical retrieval approach, in which one tries to constrain abundance and uncertainty simultaneously for a set of species. In addition, the spectral resolution is low enough that it would be challenging to uniquely distinguish molecular signatures from the spectral continuum.

For these reasons we have opted for an ``upper limit'' approach, in which we calculate the statistical significance of models including different individual molecular species, and determine the abundance at which point the molecule would be detected. The natural metric to quantify the goodness of the fit is the $\chi^{2}_{red}$ of the model fit
% Let us denote with $y_{i=1,...,n}$, $m_{i=1,...,n}$, and $\sigma_{i=1,...,n}$ respectively the observed spectrum, the model and the spectral noise for each of the $n$ spectral channels. The $\chi^{2}_{red}$ will be simply defined as:
% \begin{equation}
%    \chi^2 = \frac{1}{n-1} \sum_{i=1}^n \frac{(y_i-m_i)^2}{\sigma_i^2}
% \end{equation}
% and it will be an indication of how much, in average, the model departs from the data in units of spectral noise.
(where $\chi^{2}_{red}$ consists of the $\chi^{2}$ divided by the degrees of freedom). A $\chi^{2}_{red}$ of 1 is considered to be an optimal fit, which indicates that the model matches the data to the noise level and the information is being retrieved in its entirety (assuming correct noise estimation). A $\chi^{2}_{red}$ of less than 1 suggests that the uncertainties are overestimated, while a $\chi^{2}_{red} > 1$ indicates that either the model is incomplete or the uncertainties are underestimated. The higher above 1 that $\chi^{2}_{red}$ climbs, the less accurate the fit of the data. 

To calculate $\chi^{2}_{red}$, we utilized the optimal estimation retrieval tool in PSG. Errors in computation of normalized residuals can also lead to additional sources of uncertainty, thus PSG calculates the the standard deviation of points per residual spectrum in order to better mitigate the effect of error in $\chi^{2}_{red}$.
%In this problem, we evaluate the reduced $\chi^2$ on specific sets of models. 
Each spectral model is produced by fixing the abundance of two gases of interest to specific values, chosen in a given range; we then retrieve the planetary diameter as the only free parameter, as that can be reliably constrained by the average intensity of the observed spectral continuum (as seen in Figure 8); this also lowers our number of additional degrees of freedom to 1. We then produced two-dimensional views of the $\chi^{2}_{red}$ for each combination of gases and their respective abundances; this procedure is repeated for several combinations of gases, yielding a comprehensive view of the likelihood that the presence of those specific gases in the atmosphere may well fit the observed spectrum.

Through this process we explored the statistical parameter space to locate regions where the $\chi^{2}_{red}$ value has a minimum. Those gases characterized by a clear minimum in $\chi^{2}_{red}$ in correspondence to a specific concentration are those that are more likely to be present in the atmosphere. Those that do not yield a clear $\chi^{2}_{red}$ minimum for any concentration are likely not detectable above the noise level. 

\subsection{Results of the PSG Analysis} 

% mention all are cloud and haze free

We explored models with molecular species that are both common to planetary atmospheres and have absorption features in the WFC3 spectral band, with abundances ranging from 10$^{-7}$ to 10$^{-2}$ volume mixing ratio. CO$_{2}$ was assumed to be present in each simulation as a CO$_{2}$-only atmosphere maintained a good fit, while the second molecule varied between five others: CH$_{4}$, H$_{2}$O, H$_{2}$S, NH$_{3}$, and HCN. We chose to include any molecular species that was present at abundance greater than 1 part per billion (ppb) and had absorption lines within 200 to 1000 nm, as trace secondary molecules. We also assumed an H$_{2}$-rich atmosphere, as the combination of an H$_{2}$-rich atmosphere, trace molecules, and no clouds is the most consistent with the observations; the correlated-k opacity tables used in the forward modeling also assume an H$_{2}$-rich atmosphere.

\begin{figure*}
\centering
\gridline{\fig{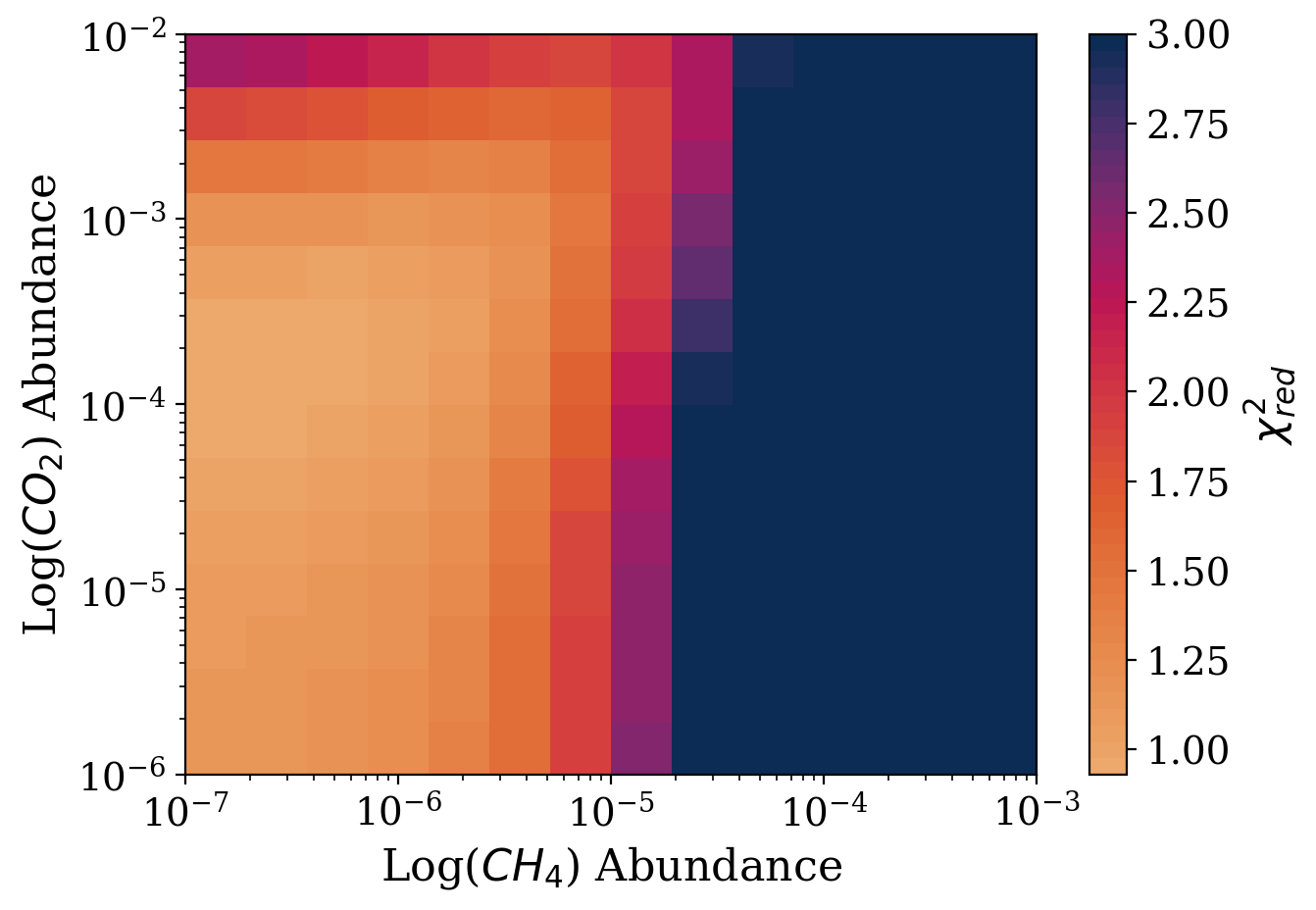}{0.334\textwidth}{}
          \fig{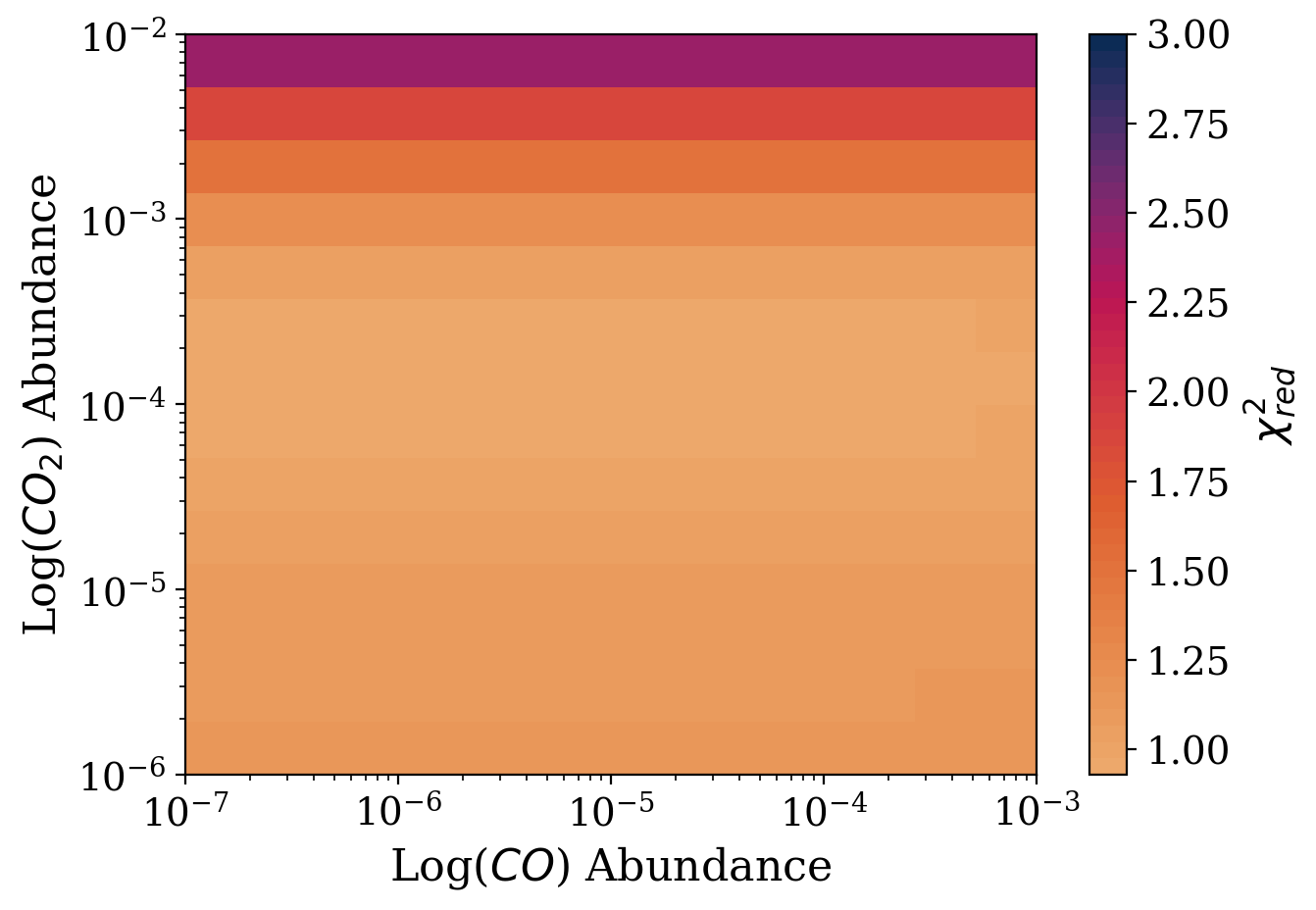}{0.334\textwidth}{}
          \fig{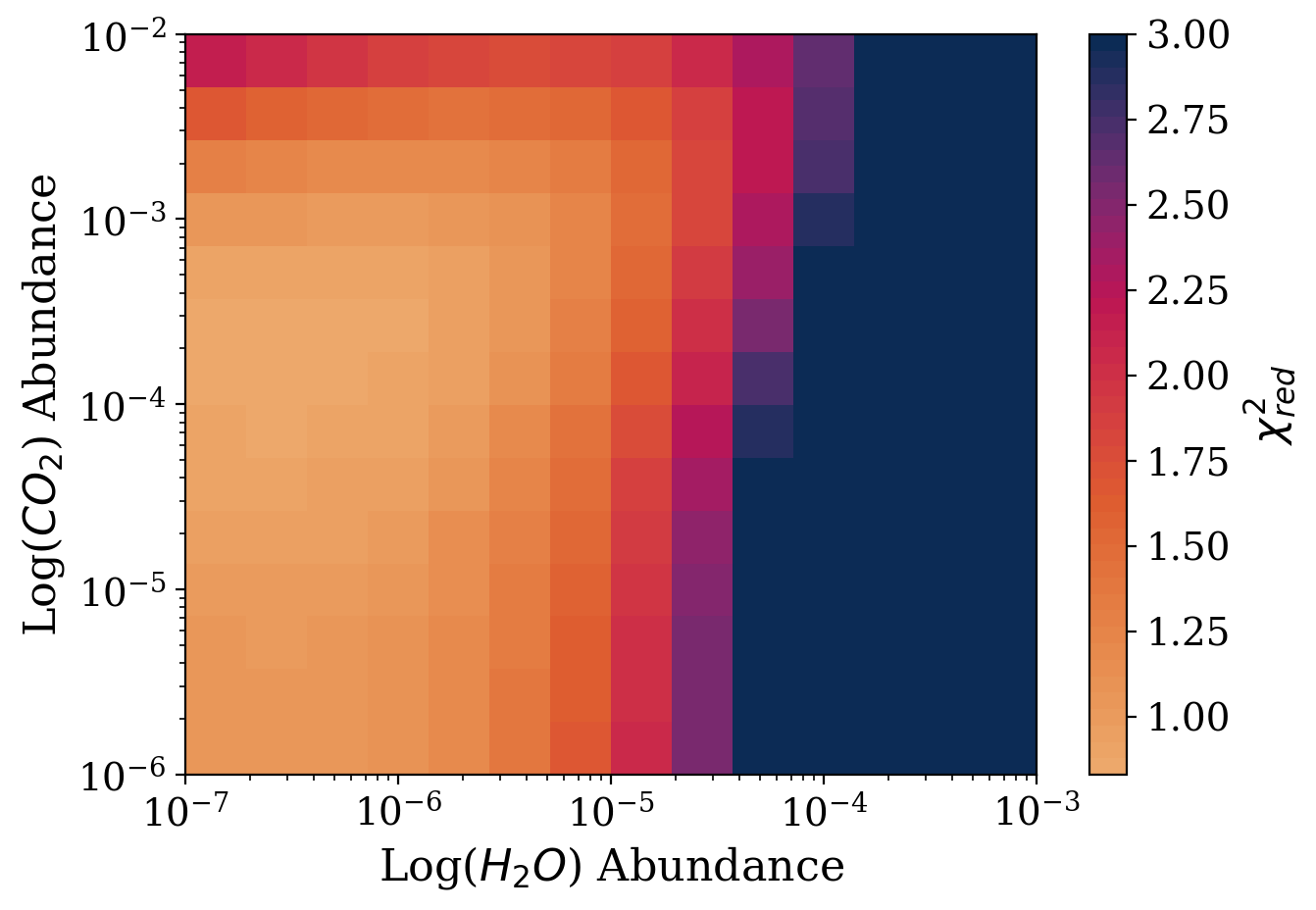}{0.334\textwidth}{}}
\vspace{-1cm}
\gridline{\fig{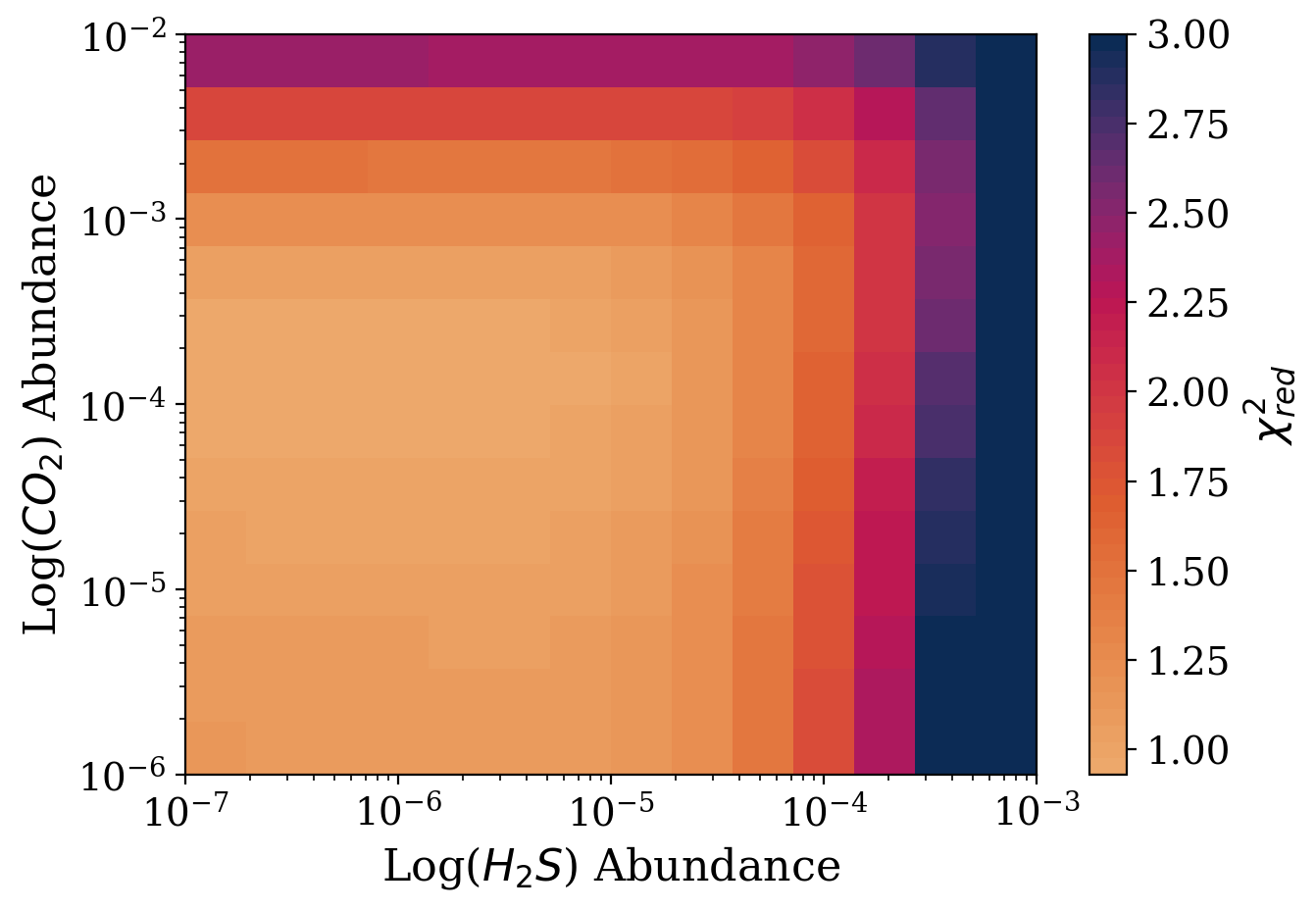}{0.334\textwidth}{}
          \fig{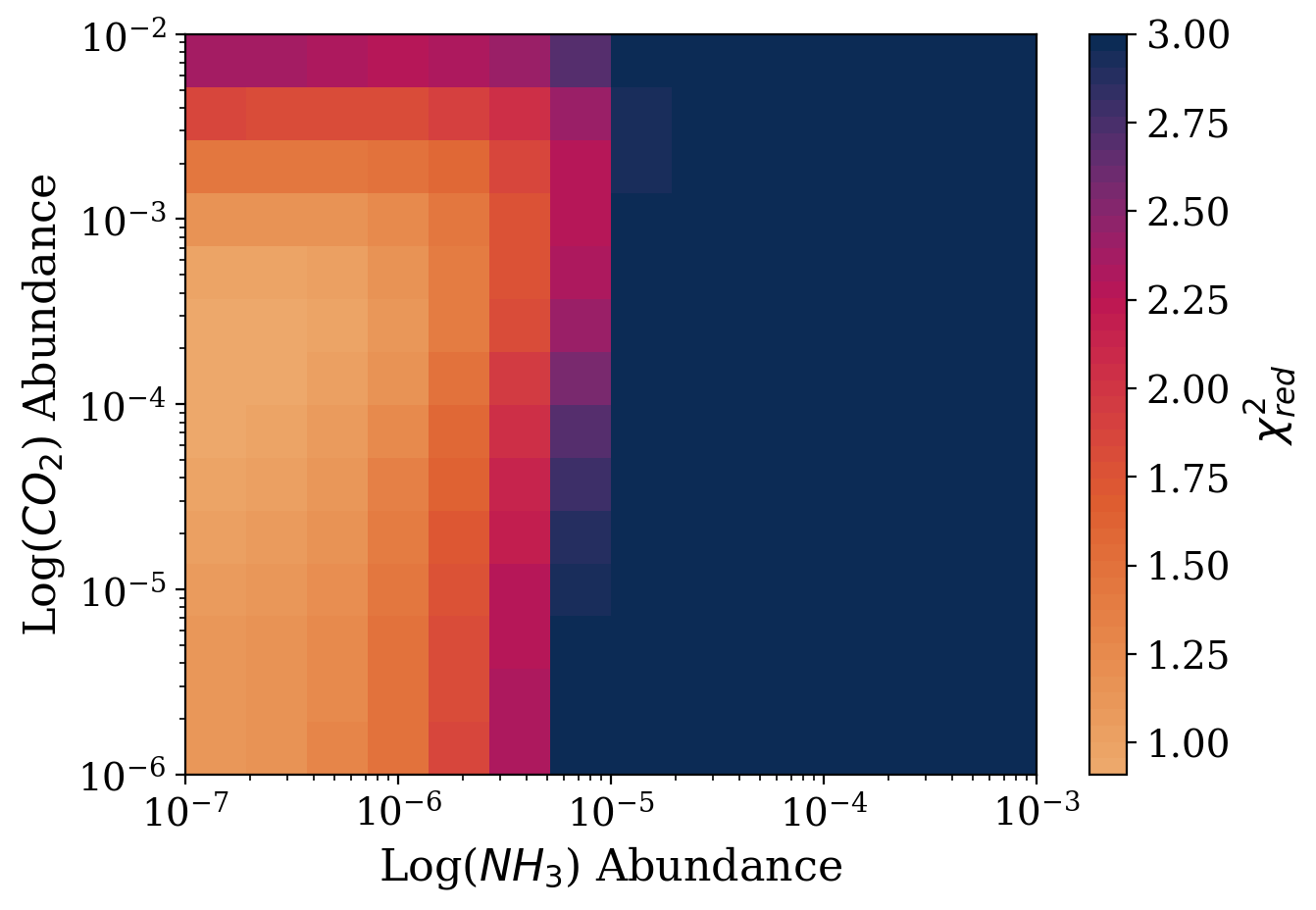}{0.334\textwidth}{}
          \fig{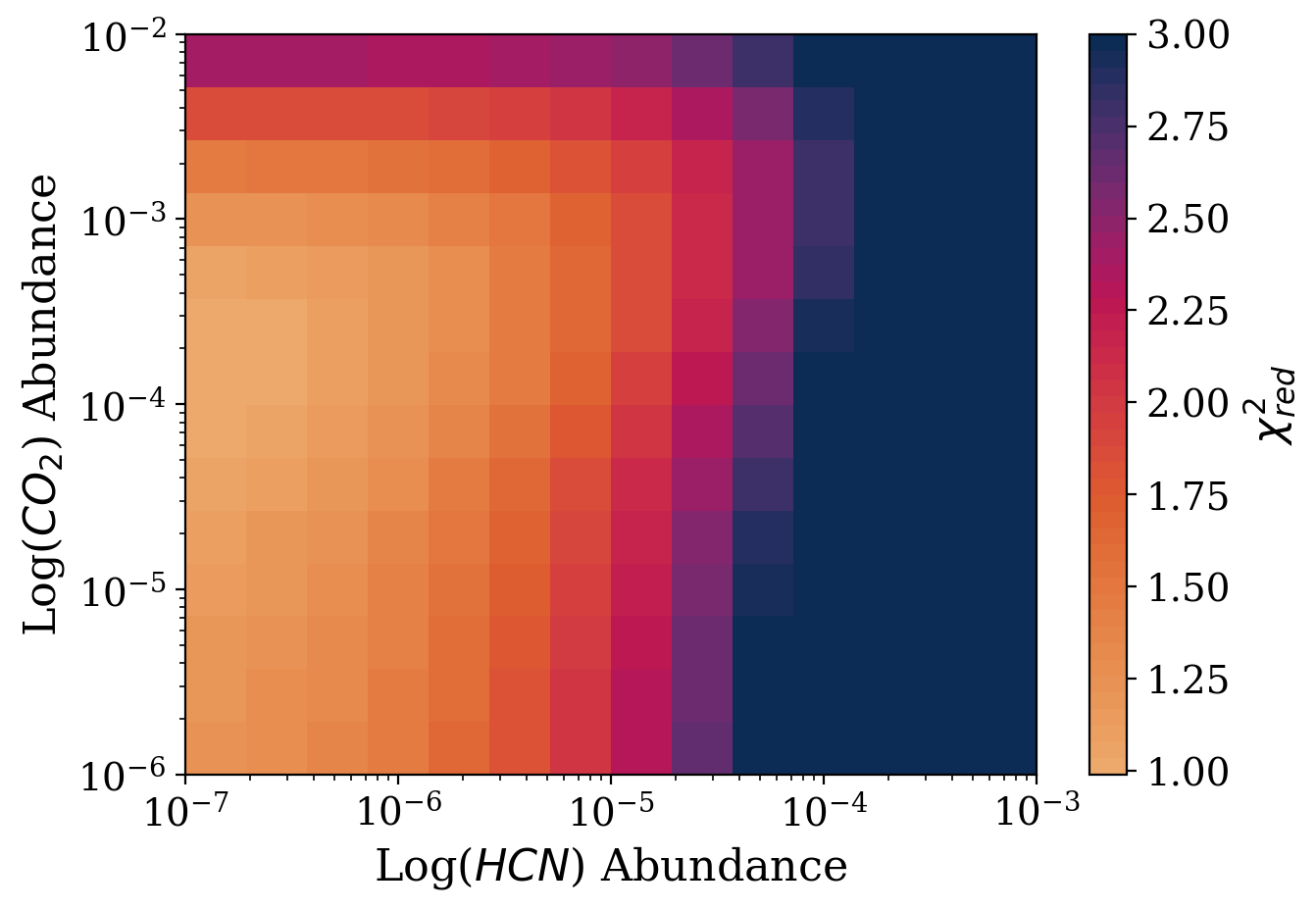}{0.334\textwidth}{}}
\vspace{-0.5cm}
\caption{The heat maps above portray the results of a two-molecule atmosphere analysis in terms of Volume Mixing Ratio (VMR). All of the presented analyses have the molecular abundance of CO$_{2}$ on the y-axis. In the top left we have CH$_{4}$, in the top middle we have CO, in the top right we have H$_{2}$O, in the lower left we have H$_{2}$S, in the lower center is NH$_{3}$, and in the lower right we have HCN. We retrieved for diameter at each abundance level, and calculated the $\chi^{2}_{red}$ to find the best fit retrieval for diameter. The best fitting models are found at 10$^{-4}$ VMR, or 10$^{2}$ ppm, CO$_{2}$ abundance and extremely low abundances, 10$^{-6}$ or lower, for other molecules.}
\label{fig:heatmaps}
\end{figure*}

We present in Figure~\ref{fig:heatmaps} the full range of $\chi^{2}_{red}$ values per molecular pair over the full range of possible abundances per molecule. As can be seen in Figure~\ref{fig:heatmaps}, the lowest $\chi^{2}_{red}$ values (between 0.5 and 1.0) can be found at approximately 10$^{-4}$ Volume Mixing Ratio (VMR), or 10$^{2}$ ppm, CO$_{2}$ abundance. For each of the other trace molecules, this minima is achieved at extremely low abundances, 10$^{-6}$ or lower. While we searched for CO, its high-energy, low-intensity bands in this wavelength range led to no fit improvement or detriment, thus it was not included in further analysis.

% \begin{figure*}
% \includegraphics[width=\textwidth]{ret_figures/heatmap_log_all.pdf}
% \caption{The heat maps above portray the results of a two molecule atmosphere analysis. All of the presented analyses have the molecule CO$_{2}$ on the y-axis. In the top left we have CH$_{4}$, in the top middle we have CO, in the top right we have H$_{2}$O, in the bottom left we have H$_{2}$S, in the bottom middle we have NH$_{3}$, and in the bottom right we have HCN. All molecules have been presented in the volume mixing ratio. We retrieved for diameter at each abundance level, and calculated the reduced $\chi^2$ to find the best fit retrieval for diameter.
% \label{fig:heatmaps}}
% \end{figure*}

\begin{figure}
\centering
\includegraphics[width=0.5\textwidth]{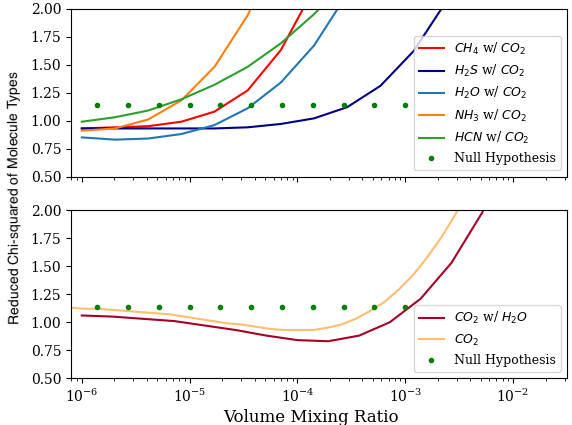}
\caption{The reduced $\chi^{2}$ when various different molecules are locked to their best-fit value are shown here. In the top panel, all CO$_{2}$ values are locked at approximately 10$^{-4}$ VMR while the other molecules are free. In the lower panel, H$_{2}$O is locked at approximately 10$^{-6}$ VMR. The value obtained with a flat transmission spectrum is presented in both panels as a dotted line, portraying the scenario in which there is no atmosphere present. This plot presents at which point every combination of molecules becomes a worse assumption than having no atmosphere - i.e. becoming an unrealistic scenario. The models in the lower panel including CO$_{2}$ provide a significantly better description of the data over a wide range of VMRs.
\label{fig:1D}}
\end{figure}

To expand on the results found in Figure~\ref{fig:heatmaps}, Figure~\ref{fig:1D} examines the direct molecular $\chi^{2}_{red}$ when different molecules are locked to a value. In the top panel of the figure, all CO$_{2}$ concentrations are locked at 10$^{-4}$ VMR (i.e. the best fit value for CO$_{2}$ according to our prior $\chi^{2}_{red}$ analysis), and the other molecules are varied across the entire VMR range of 10$^{-7}$ to 10$^{-2}$. We also plotted the fit wherein we assume there is no atmosphere present, which presents a $\chi^{2}_{red}$ of 1.14, i.e. the null hypothesis. As can be seen, the HCN and NH$_{3}$ models overtake the null hypothesis just before 10$^{-5}$ VMR, and thereafter decrease the goodness of the fit in $\chi^{2}_{red}$. The CH$_4$ model overtakes the null hypothesis just after 10$^{-5}$ VMR. These molecules, if present, have extremely low abundances. This leaves the H$_{2}$O and H$_{2}$S models, which overtake the null hypothesis line at just before and after 10$^{-4}$ VMR respectively. We extracted these molecular combinations that present the best $\chi^{2}_{red}$ values, lock the trace molecules to their respective best values, and studied them further. In the lower panel, H$_{2}$O is locked at 10$^{-6}$ VMR and we examine the scenario wherein CO$_{2}$ is the only atmospheric molecule present. When the CO$_{2}$ and H$_{2}$S combination was examined, it was seen to be almost precisely the same as the CO$_{2}$ model, and thus was omitted to prevent redundancy. Looking further at the lower panel of the plot, the CO$_{2}$ and H$_{2}$O combination results in the lowest $\chi^{2}_{red}$ value reached in our simulation, approximately 0.83. Comparatively, the CO$_2$ simulation reached a $\chi^{2}_{red}$ minima at 0.93. Both models overtake the null hypothesis just before 10$^{-3}$, at which point both models become less plausible than the potential of no atmosphere. 

\begin{figure}
\centering
\includegraphics[width=0.5\textwidth]{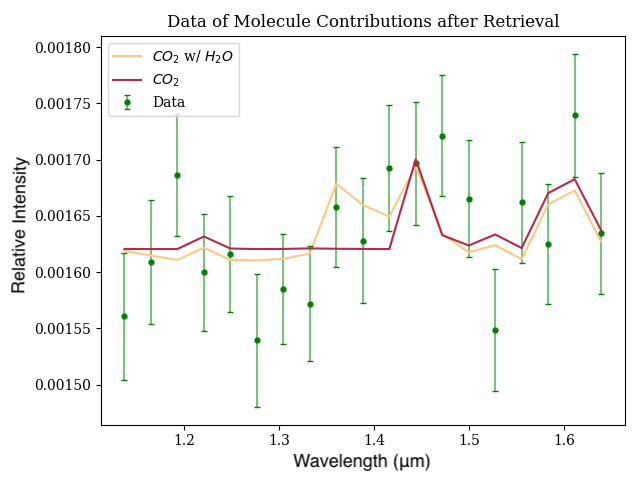}
\caption{The best retrieval results from Figure~\ref{fig:1D}'s lower panel are presented here  along with the observed data and uncertainty to portray the extent to which the model is consistent with the data.
\label{fig:ret}}
\end{figure}

Shown in Figure~\ref{fig:ret} are the retrieved models from the two best sets of abundance values described above in comparison to the L98-59~c data, with uncertainties. Figure~\ref{fig:ret} shows that our models fit the observed data well. In particular, the presence of H$_{2}$O is necessary for the model to fit the 1.35 to 1.5 $\mu$m wavelength region. 

%To further examine the results of retrieval after the null hypothesis, we turn to Figure~\ref{fig:nullhyp}. For this figure, we find the first set of values wherein all retrievals result in reduced $\chi^2$ values above the null hypothesis. The portrayed data then selects the values that correspond to the lowest reduced $\chi^2$ value that is still above the null hypothesis. In the left panel, all retrievals occur for two-molecule systems, with the listed molecule and CO$_{2}$. On the right, all retrievals occur for single-molecule systems, i.e. only the listed molecule, to see the individual contributions of each molecule. For thoroughness, we include a CO$_{2}$ only retrieval in this plot. The data is also plotted here in the lower panels with error bars, along with the H$_{2}$O retrieval data, as the model that resulted in the lowest reduced $\chi^2$ value discussed above. As can be seen in comparing the left plot, wherein the retrieval includes CO$_{2}$, with the right plot, wherein CO$_{2}$ is excluded, the model containing CO$_{2}$ fits more closely within the data error bars. The CO$_{2}$-absent model does not have the H$_{2}$O features that fit the model closer to the data, such as in the ~1.6 $\mu$m wavelength region.

\section{Discussion}
Herein, we presented HST observations of a single transit of the  planet L98-59~c. Using data from the G141 grism setting for the WFC3 instrument on HST, we extracted a spectrum using our \texttt{DEFLATE} data reduction and analysis software. The final best spectrum (i.e. $\chi^{2}_{red}$ = 0.83) shows hints of deviations from a featureless spectrum, but the uncertainties on the data are sufficiently large to be fully consistent with a flat featureless spectrum (i.e. $\chi^{2}_{red} = 1.14$). 

We performed atmospheric modeling with two different parameterized modeling schemes, using the open source tools PLATON \citep{zhang2019} and the Planetary Spectrum Generator \citep[PSG,][]{PSG}, to determine limits on composition and cloud-top pressure that would still be consistent with the data. 

The current uncertainties on the data are too large to conclusively determine anything beyond upper limits for a cloud-free or a deep-cloud ($>1 bar$) scenario. However, our analysis with the PLATON Bayesian retrieval code as well as a $\chi^{2}_{red}$ analysis using forward models from PSG are both suggestive of molecular features. The PLATON retrieval finds a best-fit result consistent with a H2-dominated atmosphere with an elevated metallicity, while the PSG analysis finds that the data is best fit with with small contributions from H$_2$O and CO$_2$, with VMRs of 5$\times10^{-6}$ and 9$\times10^{-3}$ respectively.

\subsection{Stellar Activity and Prospects for Stellar Spectral Contamination} \label{sec:contamination} 

\begin{figure}%{r}{0.6\textwidth}
     \centering 
     \includegraphics[width=0.28\textwidth]{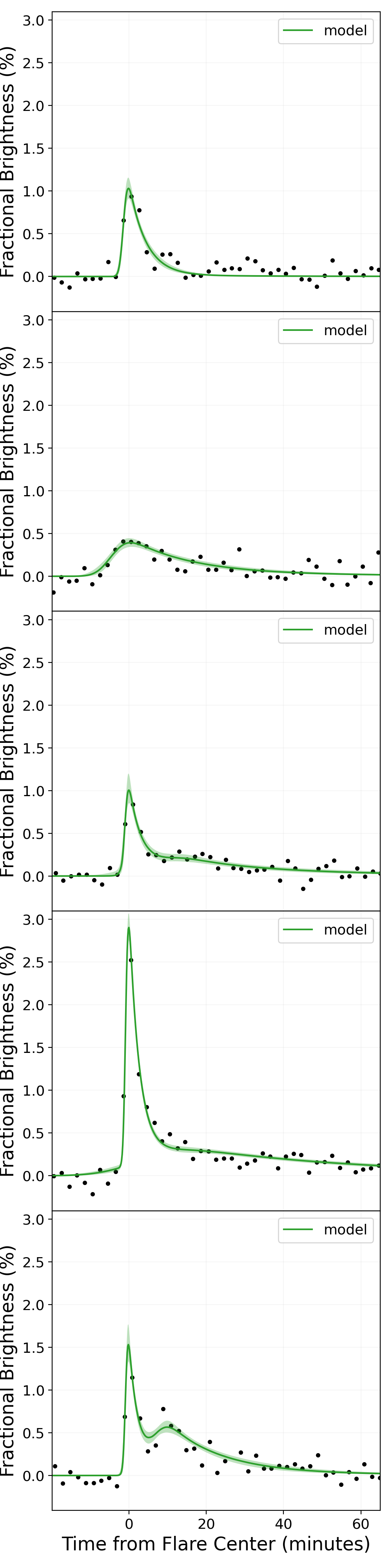}  
    \caption{Five flares of L~98-59 seen in TESS 2-minute cadence data. The flares vary in peak fractional change in stellar brightness from 0.2\% to 3.5\%. There is also one complex flare that shows multiple peaks. The green curves are flare models with uncertainty regions inferred using an MCMC method.}
    \label{fig:flares}
\end{figure}

L~98-59 was previously assumed to be a quiet star based on the analysis of TESS light curves available at the time the L~98-59 planets were discovered \citep{Kostov2019b}. However, significantly more TESS data has since been collected for this system. We reexamined the TESS 2-minute cadence data to search for signs of stellar activity. We visually inspected approximately 14 months of TESS time-series data that revealed five flares in the data, shown in Figure~\ref{fig:flares}, of which at least one appears to have multiple peaks, indicative of a complex flare profile. 

We modeled these five flares using software we developed called \texttt{xoflares} \citep{xoflares}. This version of the software implements the \citet{Mendoza2022} flare template. We sub-sampled the light curves by a factor of 7 to account for the non-linear changes in brightness occurring during a flare. The software uses Hamiltonian Monte Carlo method to efficiently sample the flare properties. We found that peak flare brightness ranged from 0.2\% to 3.5\%, which are not untypical values for mid-M dwarf stars \citep{Paudel2021,Paudel2024}.

No rotational variability is visible in TESS data, but with a rotational period of approximately 80 days \citep{Demangeon2021} low amplitude variability would be challenging to measure. While rotation modulations in photometric light curves can shed light on the presence and (lower) limits of star spots and faculae, given the flare activity identified, it is likely that L~98-59 has some level of surface inhomogeneities because flares are frequently associated with starspots \citep{Doyle2019}. 

Despite the lack of constraints on rotation modulation-and thus constraints on spot coverage on L~98-59, we performed an analysis similar to that presented in \citet{Barclay2021} to determine whether we could generate a model of stellar contamination that could mimic the transmission spectrum seen in Figure~\ref{fig:l98c_spectrum}. In \citet{Barclay2021}, HST observations of the planet K2-18~b were reanalyzed to explore whether stellar spectral contamination could be the source (or a contributing factor) of the detected planetary atmospheric signal that was reported in \citep{benneke2019b,Tsiaras2019}. For this exercise, random but plausible stellar spot models were created that emulated the observed photometric spot modulation seen in K2-18 photometric data, and simulations were performed to explore whether the water absorption signal could be a false positive arising from the inhomogeneities on the surface of the star. While the detection of water in the atmosphere of K2-18~b was not ruled out, it was determined that star spots could create the observed signal.

We performed the same exercise for L~98-59, but were not successful -- the fairly simple model used in the \citet{Barclay2021} analysis was not able to reproduce the shape of the peak around 1.45 $\mu$m that we see in the data. We could generate a peak at that location but this was always associated with a second broad peak near 1.3 $\mu$m. Moreover, generating a model that has a $>$100 ppm peak in the WFC3 G141 wavelength range that also had no clear rotational modulation in TESS data was not possible. However, the analysis performed has numerous limitations. We are cautious in saying anything definitive on the possibility that stellar surface inhomogeneities could be corrupting the observed transmission spectrum because of our poor understanding of M-dwarf surfaces and the distribution of the parameters such as spot coverage, size distribution and spectral energy distributions.

\begin{figure}%{r}{0.6\textwidth}
     \centering 
     \includegraphics[width=0.5\textwidth]{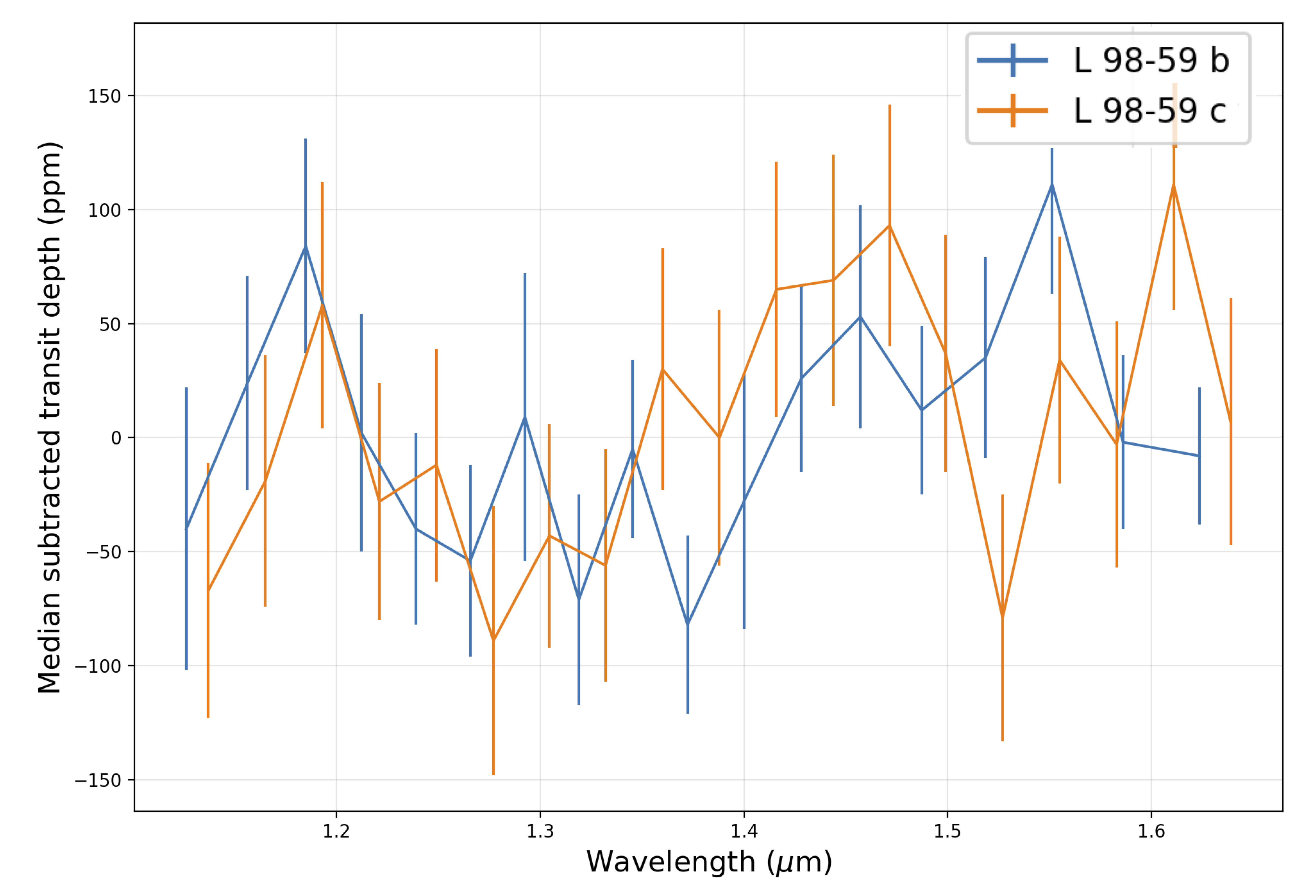}  
    \caption{Observations of L~98-59~b and c were taken by HST just 12.5 hours apart. There is correlation between the two observations which could indicate a common systematic. The L~98-59~b data were shown here are from Visit 2 of 5, and are the dataset most correlated with the HST observations of L~98-59~c.}
    \label{fig:comparebc}
\end{figure}

We also examined the HST observed transit of L~98-59~b \citep{damiano22} collected on 2020-04-07 (HST Program GO-15856, PI T. Barclay), just 12.5 hours before the transit we are examining here (see Table \ref{tab:l98_lit} for the properties of both planets). Our purpose in looking at these data was that if the L~98-59~c transit shows evidence for contamination, then the transit of planet b is likely to exhibit similar contamination signatures. This transit was the second of five observed for planet b, and showed properties largely consistent with the other four transits. However, using data from \citet{damiano22}, the standard deviation of the Visit 2 transmission spectra data was the highest of all five visits, and the transit depth measured was also the lowest of all the observed transits, albeit at only a significance of $<$2$\sigma$. Additionally, in the wavelength range between 1.2--1.4 $\mu$m the values computed for Visit 2 is either the lowest or second lowest for each bin. This results in the appearance of an absorption peak around 1.45 $\mu$m similar to that seen in the L~98-59~c transit. Figure~\ref{fig:comparebc} shows the two transmission spectra overlaid with the median subtracted from each. There are similarities in the shape of the two transmission spectra. Both spectra have dips between 1.2--1.4 $\mu$m and both have absorption peaks between 1.4--1.5 $\mu$m. 

In an effort to quantify whether the Visit 2 data is more similar to our L~98-59~c transit than the other 4 transits of L 98-59 b observed at different epochs we calculated the Pearson correlation coefficient between the planet c data and all visits of the planet b data using the \texttt{pearsonr} function from \texttt{scipy}  \citep{scipy}. Visit 2 was the most correlated with the planet c data with a probability that the correlation between the datasets occurred by chance of only 13\%, while the next highest was Visit 3 at 47\% chance of the correlation occurring by chance. This difference implies that the correlation seen in Visit 2 is 3.6x less likely to have occurred by random chance than the correlation seen for Visit 3. This is indicative of the Visit 2 data being significantly more correlated than any of the other visits. However, at less than 2-$\sigma$ confidence in this correlation, we are not in a position to say whether the correlation is owing to a common systematic between the two data sets or whether this is simply a coincidence.

%\subsection{Future Observations}
%here is where we say - but Wait! we will be getting mid cycle observations. This is what we can learn...

%JWST observations - Super great!!

\subsection{A System of Benchmark Planets}

\begin{figure}%{r}{0.6\textwidth}
     \centering    \includegraphics[width=0.5\textwidth]{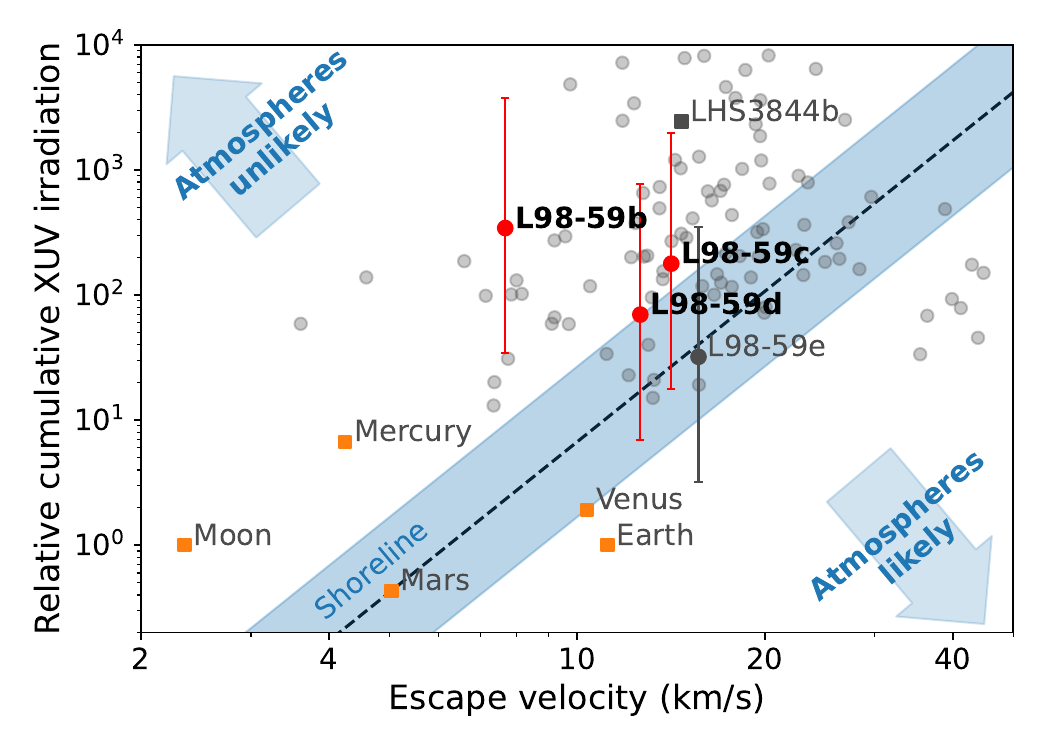}
    \caption{The L98-59 system provides a unique opportunity to probe the cosmic shoreline. In the solar system, the light blue region separates bodies that have atmospheres from those that don't \citep{Zahnle2017}. Grey dots show known rocky exoplanets with R$<$1.6 R$_\oplus$ and measured masses and radii. The integrated XUV history of L98-59 is unknown, so the vertical error bars are representative of M-dwarfs \citep{Shkolnik2014}.}
    \label{fig:shore}
\end{figure}

The L98-59 system provides an excellent opportunity to explore the atmospheres of small planets that evolved in the same stellar environment. The system also provides a unique opportunity to study the ``Cosmic Shoreline" hypothesis \citep{Zahnle2017}, which suggests that there should be a relation between planetary mass and XUV irradiation that defines a boundary between planets with and without an atmosphere. %\textbf{*plot from JWST L98-59 proposal here?*} %Currently, the evidence for this hypothesis is entirely based on our own Solar System, but the L98-59 system provides an excellent opportunity to probe the Cosmic Shoreline hypothesis. 
Although L~98-59 has been previously observed by XMM-Newton, it is challenging to use these data to determine XUV flux because (a) it is missing the UV component, and (b) using single snapshots as a measurement for the integrated XUV flux is biased because flares can occur during observations. So while the integrated XUV history of L~98-59 is not well constrained, using the uncertainties of XUV fluxes representative of M-dwarfs \citep{Shkolnik2014}, planets c and d reside near the shoreline (Figure~\ref{fig:shore}) \citep{KiteBarnett2020}. If either planet is found to retain an atmosphere, we could place a constraint on the location of the shoreline; alternatively, if the planets are found to be inconsistent with the hypothesis, it would suggest other mass loss processes (such as impact erosion) are dominant \citep{Kegerreis2020}. 

L~98-59~c and d reside in orbits that, with instellations from 4--24 times the insolation that Earth receives from the Sun \citep{Kostov2019b, Pidhorodetska2021}, place them within the ``Venus Zone'' \citep{Kane2014}, a region around a star within which a planet's atmosphere could potentially be pushed into a runaway greenhouse, producing surface conditions similar to those at Venus. Venus Zone planets are highly valuable for comparative planetology efforts that aim to characterize the conditions for planetary habitability \citep{Kane2019}. While many Venus Zone planets have been revealed \citep{Ostberg2019, Ostberg2023a}, the relatively bright L98-59 offers a unique opportunity to explore the atmospheres of this class of planets with HST and JWST \citep{Pidhorodetska2021, Ostberg2023b}. The L98-59 system is thus a benchmark system for examining Venus class planets that can help place our solar system into context.
%; in that respect, L98-59 could become a benchmark system for examining Venus-class planets and helping to shed light on why the atmospheres of Earth and Venus in our own solar system evolved to dramatically different states.
%high-mass end of the Venus thermal regime. %\textbf{*this sentence is directly from the JWST proposal, switch the language to avoid plagiarism* }

%\textbf{*Should we discuss the secondary atmosphere possibilities here or in the discussion section for future JWST observations?*}

% UNCOMMENT FIGURE FOR COSMIC SHORELINE (Who Made it?)

\section{Conclusions}
We observed a single transit of L~98-59~c with Hubble's WFC3 and find some evidence that the transmission spectrum is not flat, which could be indicative of an exoplanet atmosphere. However, the detection has low significance (2.1 $\sigma$), therefore our confidence is not sufficiently high to make this a solid claim.

Since the original discovery of the L~98-59 planetary system, more TESS data has been obtained and the star is more active than previously thought and we observe stellar flares in the star's light curve. %Based on the three flat spectra we see of L~98-59~b, one of which was collected just 14 hour prior to our L~98-59~c transit, we do not find it likely that the L~98-59~c signal is due to stellar contamination. 
We were not able to simulate any scenarios where stellar spectral contamination could cause the detected signal, although more data will be useful as the spectra of stellar surface inhomogeneities is poorly constrained. We did find some correlation between transit of L~98-59~b collected just 12.5 hours prior to our L~98-59~c transit. This transit spectrum had the most scatter of any of the five collected, and in the region of wavelength-space where contamination is most problematic, this transit spectrum provided a fairly close match the shape of the L~98-59~c spectrum -- there is only 13\% chance that the correlation between this transit of planet b and the planet c transit is due to random chance. Moreoever, this transit of planet b was much more correlated with the planet c transit than the four other planet b transits that were collected. This correlation is suggestive that both transits suffer from correlated systematics, but not conclusive as it is at a confidence level of $<$2$\sigma$.

The single transit of L 98-59 c reported herein limits the conclusions we can draw. A reliable detection will require additional data. Fortunately, L98-59~c has been selected for additional observations with HST to observe two more transits. As the observations are close to the shot noise limit, any increase in SNR should scale approximately with the square root of the number of transits observed. Furthermore, observing additional transits provides a robustness against a false positive detection owing to stellar activity.

In addition, observations from JWST (Program 1201, PI Lafreniere, using the NIRISS instrument in the Single-Object Slitless Spectroscopy mode from 0.6-2.8 $\mu$m), along with other future HST/JWST data obtained for this star, may shed light on whether the signals presented herein are from stellar spectral contamination or are astrophysical in nature and caused by a planetary atmosphere. If it is the latter, L~98-59~c could be the first planet smaller than 2 Earth-radii with a definitively-detected atmosphere.

%L 98-59 c will continue to be a benchmark target. 
%The spectrum reported here provides tantalizing hints of an atmospheric detection. However, 
%Transits of c: NIRISS Single-Object Slitless Spectroscopy (Program 1201; PI Lafreniere, GO C1, public) 0.6−2.8 μm

% Transit of d: NIRSPEC (Program 1224; PI Birkmann, GO C1, exclusive access); maybe also in Lafreniere
% NIRSPEC G395H 2.87 − 5.14 μm

%Transit of b: NIRSpec in BOTS (Bright Object Time Series) (Program 2512; PI Batalha, GO C1, public)(Program 3942, PI Damiano, GO C2, exclusive access)

% Transit of b: NIRISS SOSS and NIRSPEC BOTS (Program 4098, PI Benneke, GO C2, public)

% Star: MIRI Imaging (Program 3730; PI Diamond-Lowe, GO C2, public)

% add jwst 
% 3 programs

\acknowledgments
This work was supported by the Sellers Exoplanet Environments Collaboration (SEEC) at NASA’s Goddard Space Flight Center. This research is based on observations made with the NASA/ESA Hubble Space Telescope obtained from the Space Telescope Science Institute, which is operated by the Association of Universities for Research in Astronomy, Inc., under NASA contract NAS 5–26555. Support for program number HST-GO-15856 was provided through a grant from the STScI under NASA contract NAS5-26555. This paper includes data collected by the TESS mission. Funding for the TESS mission is provided by the NASA's Science Mission Directorate. The material is based upon work supported by NASA under award number 80GSFC21M0002. N. L. gratefully acknowledges support from an NSF GRFP.

\vspace{5mm}
\facilities{HST (WFC3), TESS}

\software{Astropy \citep{exoplanet:astropy13,exoplanet:astropy18},  
Batman \citep{kreidberg2015a},
DEFLATE \citep{sheppard2017},
KMPFIT \citep{kapetyn},
Matplotlib \citep{matplotlib},
MC$^3$ \citep{cubillos2017},
NumPy \citep{numpy},
SciPy \citep{scipy},
xoflares \citep{xoflares}
}

% \appendix

\bibliographystyle{aasjournal}
\bibliography{l98}{}

\end{document}